\documentclass[journal]{IEEEtran}
\usepackage{graphicx}
\usepackage{amsmath}
\usepackage{algorithm}
\usepackage{hyperref}
\usepackage{booktabs} % For formal tables
\usepackage{tabularx} % For adjustable-width columns

\usepackage{booktabs} % 使用三线表
\usepackage{tabularx} % 允许列宽自动调整
\usepackage{caption}  % 更好的标题控制
\usepackage{mathtools}
\usepackage{amssymb}%manthbb
\usepackage{amsmath}
\usepackage{multirow}
\usepackage{tabularx} %避免溢出
\usepackage{array}
\usepackage{makecell}
\usepackage{textgreek}
\usepackage{comment}
\usepackage{bm}
\usepackage{booktabs}
\usepackage{color}
\usepackage{tabularx}
\usepackage{algorithmicx}
\usepackage{algpseudocode}
\usepackage{subcaption}
\usepackage[normalem]{ulem}
\usepackage{xcolor}
\usepackage{colortbl} % 添加颜色支持
 
\usepackage{makecell} % 支持表格中的换行

\algnewcommand\Input{\item[\textbf{Input:}]}  % Define Input
\algnewcommand\Output{\item[\textbf{Output:}]}  % Define Output
\usepackage{graphicx} % Required for inserting images
\begin{document}

\title{Predictive Position Control for Movable Antenna Arrays in UAV Communications: A Spatio-Temporal Transformer-LSTM Framework}
%\title{Deep Learning-Driven  Adaptive Optimization and Position Prediction of Movable Antennas for UAV Communications  }
 
\author{ Kan Yu,~\IEEEmembership{Member,~IEEE}, Kaixuan Li, Xiaowu Liu, Qixun Zhang,~\IEEEmembership{Member,~IEEE}, and Zhiyong Feng,~\IEEEmembership{Senior Member,~IEEE}
 %<-this % stops a space
\thanks{This work is supported by the National Natural Science Foundation of China with Grant 62301076, Fundamental Research Funds for the Central Universities with Grant  24820232023YQTD01, National Natural Science Foundation of China with Grants 62341101 and 62321001, Beijing Municipal Natural Science Foundation with Grant L232003, and National Key Research and Development Program of China with Grant 2022YFB4300403.
}

\thanks{K. Yu is with the Key Laboratory of Universal Wireless Communications, Ministry of Education, Beijing University of Posts and Telecommunications, Beijing, 100876, P.R. China. E-mail: kanyu1108@126.com;}
\thanks{K. Li is with the School of Computer Science, Qufu Normal University, Rizhao, P.R. China. E-mail: lkx0311@126.com;}
\thanks{X. Liu (\emph{the corresponding author}) is with the School of Computer Science, Qufu Normal University, Rizhao, P.R. China. E-mail: liuxw@qfnu.edu.cn;}
\thanks{Q. Zhang is with the Key Laboratory of Universal
Wireless Communications, Ministry of Education, Beijing University of Posts and Telecommunications, Beijing, 100876, P.R. China. E-mail: zhangqixun@bupt.edu.cn;}
\thanks{Z. Feng is with the Key Laboratory of Universal Wireless Communications, Ministry of Education, Beijing University of Posts and Telecommunications, Beijing, 100876, P.R. China. E-mail: fengzy@bupt.edu.cn.}

}

% The paper headers
\markboth{IEEE Transactions on Communications,~Vol.~, No.~, 2025}%
{Shell \Baogui Huang{\textit{et al.}}: Shortest Link Scheduling Under SINR}
\maketitle
\begin{abstract}
In complex urban environments, dynamic obstacles and multipath effects lead to significant link attenuation and pervasive coverage blind spots. Conventional approaches based on large-scale fixed antenna arrays and UAV trajectory optimization struggle to balance energy efficiency, real-time adaptation, and spatial flexibility. The movable antenna (MA) technology has emerged as a promising solution, offering enhanced spatial flexibility and reduced energy consumption to overcome the bottlenecks of urban low-altitude communications. However, MA deployment faces a critical velocity mismatch between UAV mobility and mechanical repositioning latency, undermining real-time link optimization and security assurance. To overcome this, we propose a predictive MA-UAV collaborative control framework. First, optimal antenna positions are derived via secrecy rate maximization. Second, a Transformer-enhanced long short-term memory (LSTM) network predicts future MA positions by capturing spatio-temporal correlations in antenna trajectories. Extensive simulations demonstrate superior prediction accuracy (NMSE reduction exceeds 49\%) and communication reliability versus current popular benchmarks.
\end{abstract}

\IEEEpeerreviewmaketitle

\section{Introduction}\label{sec:introduction}
%城市低空通信超高可靠性与安全性需求日益迫切。复杂城市环境中，城市建筑物异构特性与多径效应导致严重的链路衰减和覆盖盲区。当前，缓解此问题的有效措施通常可以划分为两类：（1）配备大规模固定位置天线的基站优化传输功率/设计波束赋形矩阵，（2）无人机轨迹优化，（3）大规模固定天线阵列辅助的动态波束追踪，或进行三者的联合优化设计。但其固有局限性不容忽视：（1）城市建筑物复杂异构，UAV大尺度轨迹优化难以逼近性能边界，甚至优先解决的是避障问题而非通信质量保障；（2）固定天线阵列系统中，天线位置固定不变，空间自由度利用固化，无法主动规避突发性遮挡或捕捉最优信号传播路径，难以适应密集异构的城市环境高效通信。

%为了克服上述难题， movable antenna作为一种变革性解决方案，能够精细调整天线位置，通过空间自由度挖掘优化信号传输路径，联合波束赋形矩阵设计消除覆盖盲区，降低对UAV机动性/波束追踪提升通信效率的依赖，应对密集异构城市环境带来的通信挑战。然而，现有MA位置调控范式在高速移动场景下面临严峻挑战：其一，UAV与天线位置移动机械响应之间存在严重时延失配-当基站基于当前信道状态信息计算最优天线位置时，执行前UAV已移动至新位置-导致机密速率最大化策略失效，并引发危及安全的预测偏差。其二，FPA因其位置固定特点，空间自由度固化，难以主动规避遮挡或捕获信号最优传输路径。

%城市低空通信超高可靠性与安全性需求日益迫切。复杂城市环境中，动态障碍物与多径效应导致严重的链路衰减和覆盖盲区。当前，缓解此问题的有效措施通常可以划分为两类：（1）配备大规模固定位置天线的基站优化传输功率/设计波束赋形矩阵，（2）无人机轨迹优化，或联合优化（1）和（2）。但其固有局限性不容忽视：（1）UAV频繁大范围位置更新家具能耗、增加飞行控制复杂度并破坏链路稳定性，难以满足通信链路实时适配需求；（1）无人机轨迹优化需频繁执行大范围机动，不仅显著增加飞行能耗与控制复杂度，更会破坏通信链路的实时稳定性；（2）固定天线系统受限于空间自由度固化，无法主动规避突发性遮挡或捕捉最优信号传播路径，在动态环境中表现被动。
%为了克服上述难题， movable antenna作为一种变革性解决方案，基站能够精细调整天线位置，通过空间维度优化信号传输路径，提升通信质量并消除盲区，降低对UAV机动性的依赖。然而，现有MA位置调控范式在高速移动场景下面临严峻挑战：其一，UAV与天线位置移动机械响应之间存在严重时延失配-当基站基于当前信道状态信息计算最优天线位置时，执行前UAV已移动至新位置-导致机密速率最大化策略失效，并引发危及安全的预测偏差。其二，FPA因其位置固定特点，空间自由度固化，难以主动规避遮挡或捕获信号最优传输路径。
%（1）无人机轨迹优化需频繁执行大范围机动，不仅显著增加飞行能耗与控制复杂度，更会破坏通信链路的实时稳定性；（2）固定天线系统受限于空间自由度固化，无法主动规避突发性遮挡或捕捉最优信号传播路径，在动态环境中表现被动。
%为了克服上述难题，

The integration of unmanned aerial vehicles (UAVs) into low-altitude communication networks has unlocked transformative applications across urban infrastructure management and emergency response systems \cite{Geraci2022What,Baltaci2021A}. However, in complex urban environments characterized by architectural heterogeneity and dense multipath propagation, UAV-ground links experience significant signal degradation and unpredictable coverage disruptions. Conventional mitigation strategies, including fixed-antenna beamforming (BF)  \cite{Ning2023BF,He2023Full}, UAV trajectory optimization \cite{Jing2024ISAC,Li20223D}, and dynamic beam-tracking systems \cite{Ning2022A,Chen2024Beam}, exhibit fundamental limitations: (1) fixed-positioning antenna beamforming suffers from spatial-degree-of-freedom (DoF) rigidity, unable to evade dynamic blockages; (2) UAV trajectory optimization demands frequent large-scale adjustments that prioritize collision avoidance over communication quality; (3) real-time beam tracking struggles with rapidly aging channel conditions under mobility. These constraints are particularly pronounced in dense urban corridors where environmental obstacles dominate communication planning.

Movable antenna (MA) technology emerges as a paradigm-shifting solution, enabling dynamic antenna repositioning to actively exploit spatial DoF \cite{Zhu2024MAf,Zhu2025A}. Through joint optimization of antenna positioning and beamforming, MA systems mitigate coverage gaps \cite{Zhu2025A,Shao20256D}, capture optimal propagation paths \cite{Dong2024MA,Pi20256D}, and reduce trajectory-adjustment dependency \cite{Kaixuan2025}. Nevertheless, existing MA control strategies confront a critical temporal disconnect: when the base station equipped with MA arrays compute optimal antenna positions based on the current channel state information (CSI), target UAVs relocate during the inherent latency of mechanical repositioning and computational processing. This velocity mismatch causes transmissions to prior UAV locations, severely degrading link reliability. Slowing UAVs to accommodate MA response introduces unacceptable communication delays, nullifying MA’s advantages in latency-sensitive scenarios.

To bridge this gap, in this paper, we propose an intelligent MA positioning forecasting framework. Leveraging historical MA position databases and spatio-temporal correlation models, the proposed framework can predict optimal antenna configurations for future time slots—proactively compensating for CSI aging induced by high-speed UAV mobility. In particular, this work specifically addresses the pivotal question: \emph{\textbf{``How can predictive MA positioning overcome mechanical latency to guarantee reliable communications for mobile UAVs in complex urban environments?''}} The main contributions of this paper can be summarized as follows.
\begin{itemize}
     \item High-quality historical MA position dataset construction: We establish a comprehensive spatio-temporal dataset of optimal MA positions—critical for training generalizable predictors—by formulating a secrecy rate maximization problem. To address inherent non-convexity, an enhanced particle swarm optimization (PSO) algorithm generates near-optimal solutions across diverse scenarios, ensuring dataset representativeness for downstream learning;
     \item Transformer-LSTM hybrid position predictor design: A novel neural architecture integrates sequential modeling of LSTM with  attention mechanism of Transformer to forecast future antenna configurations. This synthesis minimizes the normalized mean square error (NMSE) in MA position prediction, directly enhancing communication reliability under mobility;
     \item Experimental validation in dynamic urban environments: Rigorous simulations demonstrate superior predictive accuracy and communication robustness versus benchmarks. In particular, the framework achieves $49\%$ reduction in NMSE and $14.76\%$ accuracy gain at least with a prediction length of 60 time slots, validating its efficiency for real-world deployment.
 \end{itemize}

The remainder of this paper is organized as follows. Section \ref{sec:related work} reviews the recent advancements in intelligent algorithm-driven MA technologies. In Section \ref{sec:network model}, the system model is presented, and both the secrecy rate maximization and NMSE minimization problems are formulated. Section \ref{sec:opt} details the proposed optimization and prediction framework and its efficient solution methodology. Finally, conclusions and directions for future work are provided in Section \ref{sec:conclusion}.

\section{Related Works}\label{sec:related work}
 
Featured by high spatial adaptability and the capability to dynamically adjust to time-varying environments, MAs have emerged as a promising technology for performance enhancement in wireless systems. Concurrently, extensive efforts on MA optimization have focused on the joint design of antenna positions, beamforming schemes, and transmit power, subject to diverse system constraints, to fully exploit their potential in dynamic scenarios. Subsequently, we provide a consolidated overview of recent advancements in MA position optimization technologies, as summarized in Table \ref{tab:mobility_studies}. 

\begin{table*}[t]
\caption{\small Representative Studies on Classical Methods Enabled MA}
\label{tab:mobility_studies}
\centering
\arrayrulecolor{black}
\arrayrulewidth=0.5pt
\renewcommand{\arraystretch}{1.4}
\begin{tabular}{|
>{ \centering\arraybackslash}m{0.8cm}|
>{ \centering\arraybackslash}m{4.4cm}|
>{ \centering\arraybackslash}m{1.5cm}|
>{ \centering\arraybackslash}m{0.8cm}|
>{ \centering\arraybackslash}m{1.4cm}|
>{ \centering\arraybackslash}m{1cm}|
>{ \centering\arraybackslash}m{0.8cm}|
>{ \centering\arraybackslash}m{0.7cm}|
>{ \centering\arraybackslash}m{0.7cm}|
>{ \centering\arraybackslash}m{1cm}|
}
\hline
\textbf{Rref.} & \textbf{Techniques} & \textbf{ Intelligence} & \textbf{SNR/ SINR} &
\textbf{Achievable rate/capacity} & 
\textbf{Secrecy rate} & 
\textbf{NMSE} & 
\textbf{Eve}  &
\textbf{NLoS}&
\textbf{Mobility}\\ \hline

\cite{Xiao2024Multiuser} 
& Joint design of MA positions, receive combining matrix, and transmit power
& $\times$ 
& $\checkmark$ 
& $\checkmark$
& $\times$ 
& $\times$ 
& $\times$ 
& $\checkmark$
& $\times$ 
\\  \hline

\cite{Shao20256D} 
& Design of 6DMA deployment
& $\times$  
& $\checkmark$ 
& $\checkmark$  
& $\times$ 
& $\times$ 
& $\times$ 
& $\times$
& $\times$\\ \hline

\cite{Mei2024Movable}
& Design of MA positions
& $\times$
& $\checkmark$ 
& $\checkmark$ 
& $\checkmark$ 
& $\times$ 
& $\checkmark$ 
& $\times$ 
& $\times$ \\  \hline

\cite{Kaixuan2024}
& Joint design of MA positions and BF
& $\times$
& $\checkmark$ 
& $\checkmark$ 
& $\checkmark$ 
& $\times$ 
& $\checkmark$ 
& $\times$ 
& $\checkmark$ 
\\  \hline

\cite{Yujia2024}
& Joint design of MA positions and BF
& $\times$
& $\checkmark$ 
& $\checkmark$ 
& $\checkmark$ 
& $\times$ 
& $\checkmark$ 
& $\checkmark$
& $\checkmark$\\  \hline

\cite{Tang2025Deep} 
& Joint design of MA positions and BF 
& $\checkmark$
& $\checkmark$ 
& $\times$ 
& $\times$ 
& $\times$ 
& $\times$
& $\times$ 
& $\times$ \\  \hline

\cite{Weng2024Learning} 
& Joint design of MA positions and BF
& $\checkmark$
& $\checkmark$ 
& $\checkmark$ 
& $\times$ 
& $\times$ 
& $\times$ 
& $\times$  
& $\checkmark$ \\  \hline

\cite{Xie2025A} 
& Joint design of MA positions and BF
& $\checkmark$
& $\checkmark$ 
& $\checkmark$ 
& $\times$ 
& $\times$ 
& $\times$
& $\times$ 
& $\checkmark$ \\  \hline

\cite{Shao2025Hybrid} 
& Joint design of MA positions and BF
& $\checkmark$
& $\checkmark$ 
& $\checkmark$ 
& $\times$ 
& $\times$ 
& $\times$
& $\times$
& $\checkmark$\\  \hline

\cite{Zhao2025Movable} 
& Joint design of MA positions, BF, user selections, and transmit power
& $\checkmark$
& $\checkmark$ 
& $\checkmark$ 
& $\times$ & $\times$ 
& $\times$ & $\times$
& $\checkmark$\\  \hline

\cite{Bai2024Movable} 
& Joint design of UAV’s trajectory and the MA’s orientation  
& $\checkmark$ 
& $\checkmark$ 
& $\checkmark$ 
& $\times$
& $\times$ 
& $\times$ 
& $\checkmark$
& $\checkmark$\\  \hline

\cite{Jang2025New} 
& Joint design of MA positions and the channel angle estimation
&$\checkmark$
&$\times$  
&$\times$  & $\times$ 
& $\checkmark$
& $\times$ 
& $\times$
&$\checkmark$\\  \hline

\textbf{our work} 
& Prediction of MA positions 
& $\checkmark$ 
& $\checkmark$ 
& $\checkmark$ 
& $\checkmark$  
& $\checkmark$  
& $\checkmark$  
& $\checkmark$
& $\checkmark$\\  \hline

\end{tabular}
\vspace{0.5cm}
\raggedright
\end{table*}

\subsection{Traditional algorithm-based MA technologies}
Various optimization techniques, such as PSO, successive convex proximation (SCA), and projected gradient ascent (PGA), have been widely adopted for MA positions optimization, aiming to fully exploit the spatial adaptability of MA for enhanced communication performance and security.   

To validate the advantage of MA in enhancing communication performance, \emph{Xiao et al.} formulated a joint optimization problem involving MA positions, receive combining matrix, and users’ transmit power to maximize the minimum achievable rate among multiple users, which was efficiently solved using a two-loop iterative algorithm based on PSO and a low-complexity alternating optimization (AO) method \cite{Xiao2024Multiuser}.
Furthermore, in \cite{Shao20256D}, a six-dimensional movable antenna (6DMA) configuration was developed, integrating 3D positional and 3D orientational control. By employing a Monte Carlo simulation-assisted AO method for antenna position adjustment under practical deployment constraints, the proposed design achieved greater spatial flexibility and further enhanced communication performance.
In \cite{Wang2024Movable}, \emph{Wang et al.}  introduced an effective iterative optimization strategy, where the MA positions are updated through successive convex approximation (SCA), while the transmit beamforming is refined using a second-order cone programming (SOCP) formulation.
Nevertheless, the aforementioned studies overlook the impact of potential Eve, which lead to severe information leakage and undermine the confidentiality of legitimate communications.
Considering the communication security, in \cite{Mei2024Movable}, \emph{Mei rt al.} formulated a secrecy rate maximization problem by jointly designing the MA positions over the discrete sampling points and the transmit BF.A partial enumeration algorithm was proposed to obtain its optimal solution without the need for high-complexity exhaustive search.

Considering the mobility of transmitters or receivers in practical scenarios, in \cite{Kaixuan2025}, we investigated a dynamic communication setting in which the UAV serves as an aerial BS equipped with MA, aiming to explore the performance differences between UAV-based macroscopic movement and MA-based microscopic movement through secrecy rate analysis.
In another work \cite{Yujia2024},  taking into account the characteristics of NLoS links, we investigated secure communications in a multi-eavesdropper scenario, where the optimal beamforming and MA positions were jointly obtained using the PGA and simulated annealing (SA) methods.

\subsection{Intelligent algorithm-driven MA technologies}
Driven by the need for flexible and adaptive optimization in MA-enabled wireless systems, intelligent algorithms have recently been explored as a promising approach to addressing challenges such as dynamic environments, imperfect CSI, and computational constraints.

Given the potential of MAs in anti-jamming communications, \emph{Tang et al.} formulated a signal-to-interference-plus-noise ratio (SINR) maximization problem, where receive beamforming is solved via the Rayleigh quotient in \cite{Tang2025Deep}. In addition, a neural network architecture based on a multilayer perceptron (MLP) was designed to optimize antenna positioning, with network parameters trained using stochastic gradient descent. This approach enables offline training and supports online inference with marginal computational complexity. 
Focusing on MA-enabled multi-receiver communication systems, \emph{Weng et al.} proposed a heterogeneous multi-agent deep deterministic policy gradient (MADDPG) algorithm, in which two agents independently learn the beamforming and mobility strategies of the MAs, thereby enabling offline learning in imperfect CSI scenarios. While the joint optimization of beamforming and antenna positioning can significantly enhance system performance, it inevitably introduces computational complexity that may impact overall efficiency \cite{Weng2024Learning}.  
To address these challenges, a novel heterogeneou MADDPG framework was proposed by \emph{Xie et al.} to  optimization the position of MAs in \cite{Xie2025A}, in which heterogeneous agents independently adjust the antenna configuration. This architecture effectively enhances sensing accuracy, communication reliability, and power transfer efficiency, thereby strengthening the system’s capabilities and adaptability to dynamic environments. 
In \cite{Shao2025Hybrid}, \emph{Shao et al.} proposed an efficient hybrid-field generalized 6DMA channel model, and leveraged the low-complexity design capabilities of deep reinforcement learning (DRL) to jointly optimize the position, orientation, and beamforming of the 6DMA. 
In response to the latency and energy constraints associated with large language model (LLM) training over bandwidth-limited wireless links in 6G, \emph{Zhao et al.} proposed an optimization framework that jointly adjusts the number of global rounds, CPU frequency, mini-batch size, MA positions, and beamforming strategy, aiming to address the delay, energy, and communication challenges inherent in federated fine-tuning of LLMs, thereby enhancing model performance and training efficiency \cite{Zhao2025Movable}. 
Considering a scenario where a UAV is equipped with MAs, in \cite{Bai2024Movable}, \emph{Bai et al.} designed  a joint optimization framework to minimize the total data collection time by simultaneously optimizing the UAV’s trajectory and the orientation of the MAs. To further reduce computational complexity, a DRL-based strategy was developed, in which the agent's observation space is simplified using the azimuth angles and distances between the UAV and each backscatter device. 
In \cite{Jang2025New}, \emph{Jang} and \emph{Lee} proposed a learning-assisted channel estimation framework that jointly models the position adjustment of MAs and the channel estimation function. The evaluation results, measured by the NMSE, demonstrate a substantial reduction in estimation error, highlighting the effectiveness of the proposed framework in accurately reconstructing channel states for MA-enabled systems.

However, in contrast to previous works that mainly focus on optimizing antenna positions, our study explicitly addresses the mismatch between the movement speed of the MA and that of the users by incorporating antenna position prediction into the optimization framework. This approach enables the system to proactively adapt to user mobility, thereby enhancing the practicality and robustness of movable antenna deployment in dynamic wireless environments.

Notations: In this paper, $ (\cdot)^{\rm \mathbf{H}}$ and  $ (\cdot)^{\rm \mathbf{T}}$ denote the Hermitian (conjugate transpose) and transpose operations, respectively.

\begin{figure}
\centering
\includegraphics[width=3in]{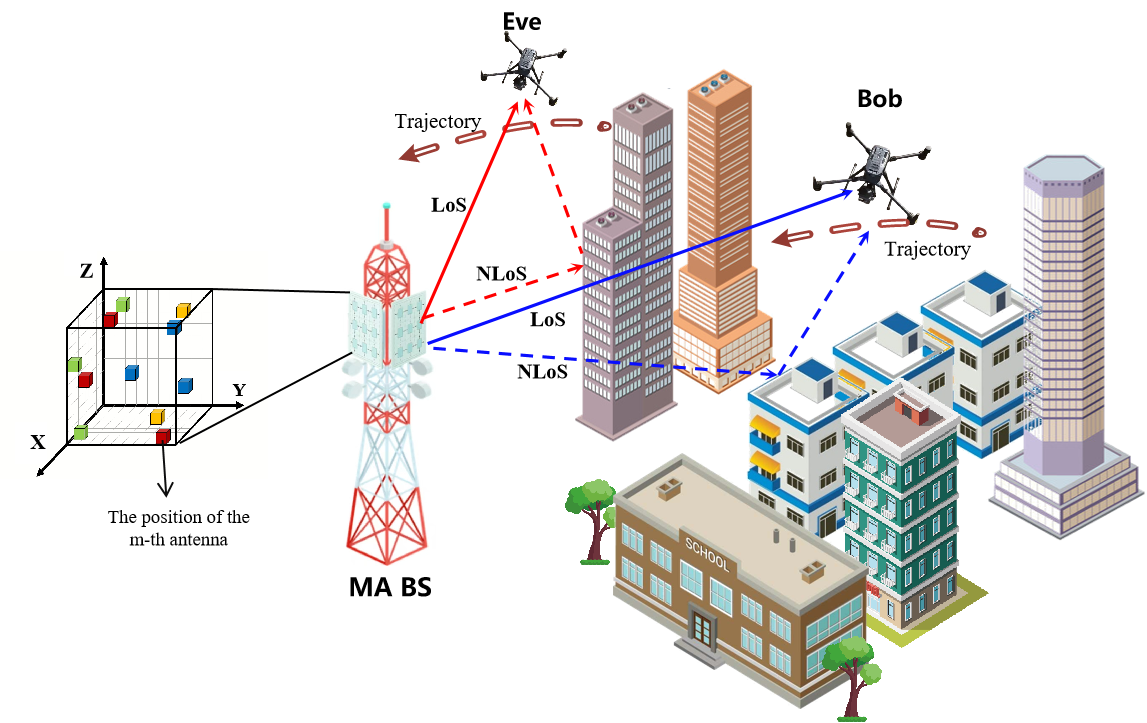}
\caption{\small Network model}
\label{fig:scenario}
\end{figure}

\section{Network Model and Preliminaries}\label{sec:network model}
As depicted in Fig. \ref{fig:scenario}, a dynamic BS-to-UAV communication scenario is considered, where the line-of-sight (LoS) and non-line-of-sight (NLoS) conditions fluctuate due to UAV mobility and potential environmental factors. Specifically, the BS, equipped with $M$ MAs, transmits secret information to a legitimate UAV, referred to as Bob, which is equipped with a single receiving antenna. Meanwhile, an aerial eavesdropping UAV (Eve), also equipped with a single antenna, attempts to intercept the transmitted message. The BS is fixed at position $(0, 0, H)$, where $H$ denotes its deployment altitude. The Bob and Eve fly along a designated 3D trajectory, and their positions at time slot $t$ are denoted by $(x_b[t], y_b[t], z_b[t])$ and $(x_e[t], y_e[t], z_e[t])$, respectively. Each element in the MA array is capable of adjusting its 3D positions, and the position of the $m$-th antenna at time $t$ is denoted as $\textbf{p}_m[t] = (x_m[t], y_m[t], z_m[t])$ in the Cartesian coordinate system.

\subsection{Communication model}

Considering the multipath effects in practical environments, the received signal at the Bob/Eve is modeled as comprising one LoS  path with $L_t^{\rm los}$ and $L_t^{\rm nlos}$ NLoS paths. Consequently, the total number of transmission paths satisfies $L_t =L_t^{\rm los}+L_t^{\rm nlos}$. The channel coefficients are modeled using a Rician distribution \cite{Ricain1,Ricain2}, which effectively captures the fading characteristics under mixed LoS/NLoS propagation conditions
\begin{equation}
\begin{aligned}
y_i &=\sqrt{ \frac{\kappa}{\kappa+1}}\underbrace{\mathbf{h}_i^{\rm los}[t]^{\rm \mathbf{H}} \mathbf{w} s[t]}_{\text{LoS link}} 
+ \sqrt{ \frac{1}{\kappa+1}}\underbrace{\mathbf{h}_i^{\rm nlos}[t]^{\rm \mathbf{H}} \mathbf{w} s[t]}_{\text{NLoS link}}  + \omega_i^2[t] \notag\\ 
&=\mathbf{H}_i[t]\mathbf{w} s[t]+ \omega_i^2[t],~~\text{$i\in\rm\{Bob, Eve\}$}
\end{aligned}
\end{equation}
where $\kappa$ denotes the Rician factor defined as the ratio of the power of the dominant LoS path to the total power of the scattered NLoS components, $\mathbf{h}_i^{\rm los}[t] \in \mathbb{C}^{M \times 1}$ represents the LoS channel vector at the $t$-th time slot, which is dependent on the instantaneous positions of both the BS and the Bob/Eve, $\mathbf{h}_i^{\rm nlos}[t] \in \mathbb{C}^{M \times 1}$ denotes the random NLoS channel component, $\mathbf{w}=\sqrt{ \frac{P_{\rm com}}{M}} 1_{\rm M} \in \mathbb{C}^{M \times 1}$ is the fixed equal-power transmit vector; $1_{\rm M}$ is the all-ones column vector of length $M$, $P_{\rm com}$ denotes the communication power, $s[t]$ is the transmit signal with zero mean and unit power, and $\omega_i^2[t]$ denotes the additive noise at the Bob/Eve.  

The spatial dependence of the channel vector on the antenna positions is characterized by representing the direction of the $j$-th transmission path $(1 \le j \le L_t)$ as a normalized spatial vector
\begin{equation}
\mathbf{r}_j = 
\begin{bmatrix}
\alpha_{j} \\
\beta_{j} \\
\gamma_{j}
\end{bmatrix}
=
\begin{bmatrix}
\cos\theta_{j} \cos\phi_{j} \\
\cos\theta_{j} \sin\phi_{j} \\
\sin\theta_{j}
\end{bmatrix},
\end{equation}
where the elevation angle $\theta_{j} $ and the azimuth angle $\phi_{j}$ are defined according to the relative orientation of the transceivers. Each component is obtained via trigonometric relations, capturing the directional characteristics. For the NLoS link, the angles are generated randomly, whereas for the LoS link, they can be directly determined based on the positions of the BS and user $i$.
Accordingly, the steering vector from the $m$-th MA to user $i$ at $t$-th time slot is constructed by
\begin{equation}
\mathbf{g}_i(\textbf{p}_m)[t] = 
\begin{bmatrix}
e^{j\frac{2\pi}{\lambda} \textbf{p}_m[t]^{\rm \mathbf{T}}\textbf{r}_1[t]},
 \cdots ,
e^{j\frac{2\pi}{\lambda} \textbf{p}_m[t]^{\rm \mathbf{T}}\textbf{r}_{l_t}[t]}
\end{bmatrix}^{\rm \mathbf{T}}, 
\end{equation}
where $l_t \in \{ L_t^{\rm los},L_t^{\rm nlos} \}$, $\lambda $ is the carrier wavelength, $\frac{2\pi}{\lambda} \mathbf{p}_m^{\rm \mathbf{T}} \mathbf{r}_j$ captures the relative phase shift induced by the position of the $m$-th antenna along the $j$-th propagation path.
Therefore, the transmit field response matrix from BS to Bob/Eve is  
\begin{equation}
\mathbf{G}_{i} = \left[ \mathbf{g}_i(\textbf{p}_1), \ldots, \mathbf{g}_i(\textbf{p}_m), \ldots, \mathbf{g}_i(\textbf{p}_M) \right] \in \mathbb{C}^{l_t \times M}.
\end{equation}

The antennas at both Bob and Eve are fixed to capture the received signals, and their received field response vectors are
\begin{equation}
\mathbf{f} = {\left[ 1,1,\ldots,1\right]}^{\rm \mathbf{T}} \in \mathbb{R}^{l_t \times 1}.
\end{equation}
To capture the multipath propagation characteristics introduced by the MAs, the path response matrix of the NLoS link for each transmission path between the MAs and Bob/Eve is defined as $\boldsymbol{\Sigma}_i^{\rm nlos} = \text{diag}(\sigma_{i,1}, \sigma_{i,2}, \ldots, \sigma_{i,L_t^{\rm nlos}}) \in \mathbb{C}^{L_t^{\rm nlos} \times L_t^{\rm nlos}}$ \cite{Zhou2024MA}, where $\boldsymbol{\Sigma}_i^{\rm nlos}$ is a diagonal matrix, with diagonal elements being mutually independent and conforming to the same distribution $\mathcal{CN}\left(0,\, \beta_0 d_i^{-\alpha} {L_t^{\rm nlos}}^{-1}\right)$, $\beta_0 $ signifies the reference path loss at a standardized distance of 1 meter, $\alpha$ is the path loss exponent and $d_i$ is the distance between the BS and Bob/Eve.
For the LoS link, $\boldsymbol{\Sigma}_i^{\text{\rm los}} = \beta_0 d_i^{-\alpha}$.

Building upon the spatial and multipath channel modeling described above, the field response channel vector from the BS to the Bob/Eve is expressed as \cite{Zhu2024MA}
\begin{equation}
\mathbf{h}_i= \left( (\mathbf{f}_i)^{\rm \mathbf{T}} \boldsymbol{\Sigma}_i \mathbf{G}_i \right)^{\rm \mathbf{T}} \in \mathbb{C}^{M \times 1}.
\end{equation}
To quantitatively evaluate the communication quality, the signal-to-noise ratio (SNR) at the Bob/Eve can be defined as the ratio of the received signal strength to the noise. Then, the instantaneous SNR at time slot $t$ is expressed as
\begin{equation} \label{eq:SNR_b}
\mathrm{\gamma }_i[t] = \frac{\left| \mathbf{H}_i^{\rm \mathbf{H}}[t] \mathbf{w} \right|^2}{\omega_i^2[t]}
\end{equation}

\subsection{Performance Metrics of MA Positions Forecasting}
%为什么采用这两个作为度量指标
In the context of physical layer security, MA technology dynamically adjusts antenna positions to compensate for signal energy attenuation, exerting distinct effects on the legitimate link and the eavesdropping link, respectively. To predicate future antenna positions in the future time slots more exactly, a high-quality antenna positions database in current time slots is need. As a result, we aim to maximize the secrecy rate to construct the database. 
%In particular, the secrecy rate quantifies the achievable rate at which information can be securely transmitted to a legitimate receiver while preventing potential Eves from recovering the transmitted content, thereby directly reflecting the effectiveness of physical-layer security mechanisms.

%To evaluate the communication performance at each time slot and generate a high-quality dataset for subsequent prediction tasks, the secrecy rate is adopted as the primary performance metric.
\subsubsection{Secrecy rate}
The secrecy rate quantifies the achievable rate at which information can be securely transmitted to a legitimate receiver while preventing potential eavesdroppers from recovering the transmitted message, thereby directly reflecting the effectiveness of physical-layer security mechanisms. Based on the result of \cite{Mert2021}, the achievable rate of the Bob/Eve is
\begin{equation}
\tau[t] = \left[ R_b[t] - R_e[t] \right]^+, \\
\end{equation}
where $[ a]^+ \triangleq \max \{a,0\} $ \cite{Hu2024MA}. 

To evaluate the accuracy of antenna position prediction, the NMSE is employed as a scale-invariant metric that measures the relative discrepancy between the predicted and actual antenna positions \cite{LLM4P,Jang2025New}.
\subsubsection{ NMSE}
A lower NMSE directly translates into more precise antenna position estimation, which is critical for ensuring reliable communications. Mathematically, it is defined as
\begin{equation}
{\rm NMSE} = \frac{\left\| \widehat{\mathbf{y}} - \mathbf{y}\right\|_{\rm F}^2 }{\left\| \mathbf{y}\right\|_{\rm F}^2 },
\end{equation}
where $\widehat{\mathbf{y}}$ and $\textbf{y}$ denote the predicted and true values.

\subsection{Optimization Problem Formulation}
Motivated by anticipatory antenna control principles, our framework shifts from reactive antenna repositioning to proactive forecasting of future optimal positions based on historical data, which jointly satisfies antenna mobility constraints and UAV dynamic adaptation, enhancing communication reliability. The proposed framework comprises two phases: optimal antenna position database construction and predictive position modeling.

\subsubsection{Optimal antenna positions dataset construction} Let $T_{\rm hist}$ denote the last time slot index of the available historical antenna positions. Given $t \in \{ 1, ...,T_{\rm hist}  \}$, the optimal antenna positions, denoted as $\mathbf{P}_\mathrm{MA}[t]$, are obtained by solving a secrecy rate maximization problem as follows:
\begin{subequations} \label{eq:P1}
\begin{align}
\text{P1:} \quad
&\max_{\mathbf{P}_{\rm MA}} \quad  \tau[t] \\
\mbox{s.t.}\quad 
& \left| \mathbf{P}_m[t] - \mathbf{P}_m[t-1] \right| \leq d_{\max}^{\rm slot}, \forall m \in \{1, \dots, M\}, \notag\\ 
&\forall t>0 \label{eq:P1-sub1} \\
& \textbf{P}^{\min}_m \in {\left[ x^{\min}_{m},x^{\max}_{m} \right]} \times {\left[ y^{\min}_{m},y^{\max}_{m} \right]} \times {\left[ z^{\min}_{m},z^{\max}_{m} \right]},  \label{eq:P1-sub2}
\end{align}
\end{subequations}
where constraint~\eqref{eq:P1-sub1} restricts the movement of the $m$-th antenna element between adjacent time slots, ensuring that its displacement does not exceed the maximum permitted step size $d_{\max}^{\rm slot}$. Constraint~\eqref{eq:P1-sub2} guarantees that the antenna’s three-dimensional position remains within the specified feasible region. The optimal antenna positions are difficult to obtain, since the Problem \eqref{eq:P1} is non-convex and there exists strong coupling among the antenna elements. In the following subsection, we propose a solution based on the particle swarm optimization (PSO) algorithm.

\subsubsection{Antenna Position prediction} Utilizing the dataset constructed from the historical antenna positions, the NMSE between the predicted and the actual optimal positions is adopted to quantitatively assess the prediction accuracy. Therefore, the MSE minimization problem is formulated as \cite{LLM4P}
\begin{subequations} \label{eq:P2}
\begin{align}
&\text{P2:}
\min_{\mathbf{P}_\mathrm{MA}} \quad\mathbb{E} \left[   
\frac{ \sum_{t_{\rm pre}=1}^{T_{\rm pre}}\left\| \widehat{\mathbf{P}}_\mathrm{MA}^{t_{\rm pre}} - \mathbf{P}_\mathrm{MA}^{t_{\rm pre}} \right\|_{\rm F}^2 }{ \sum_{t=1}^{T_{\rm pre}}\left\| \mathbf{P}_\mathrm{MA}^{t_{\rm pre}} \right\|_{\rm F}^2 }
\right] \\
&\mbox{s.t.}\quad  \left( \widehat{\mathbf{P}}_\mathrm{MA}^{1}, \ldots, \widehat{\mathbf{P}}_\mathrm{MA}^{T_{\rm pre}} \right)
= f \left( \mathbf{P}_\mathrm{MA}^{T_{\rm hist}}, \ldots, \mathbf{P}_\mathrm{MA}^{t_{\rm hist}} \right)
&  \label{eq:P1-sub3}
\end{align}
\end{subequations}
where $T_{\rm pre}$ denotes the number of future time slots to be predicted; $\widehat{\mathbf{P}}_{\mathrm{MA}}^{t}$ and $\mathbf{P}_{\mathrm{MA}}^{t}$ represent the predicted and true antenna positions at time slot $t$, respectively, $f(\cdot)$ denotes the nonlinear mapping that infers future antenna positions based on historical spatial trajectories. The problem of minimizing the NMSE is addressed using the deep learning method. A detailed description of the algorithm, including its procedure and steps, can be found in Section \ref{sec:opt}.

\section{Performance Analysis and Secret Rate Maximization}\label{sec:opt}
\begin{figure*}
\centering
\includegraphics[width=6.5in]{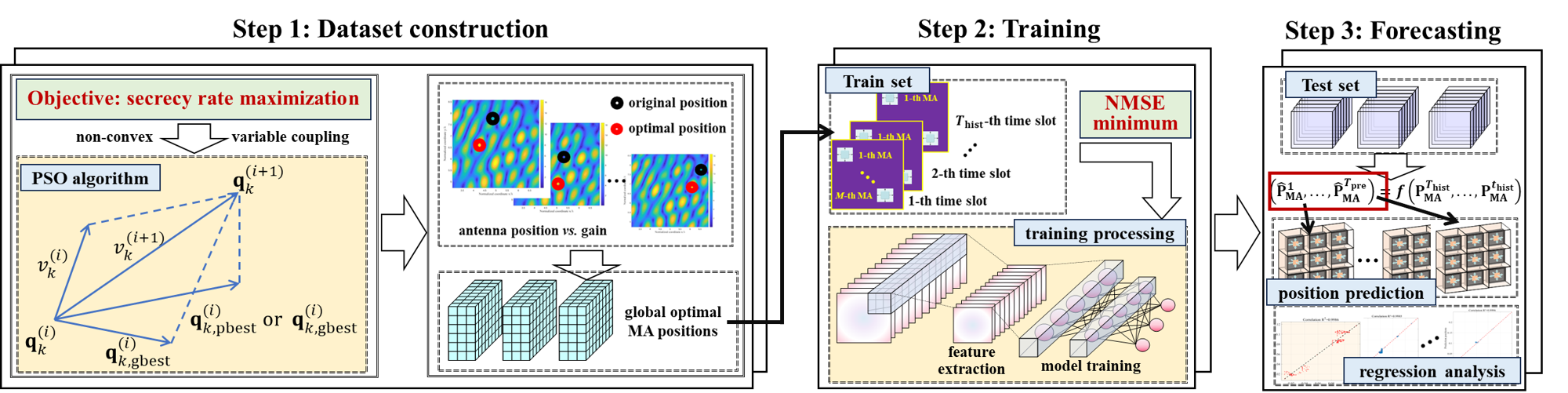}
\caption{\small The framework of the proposed antenna position optimization and prediction system}
\label{fig:process}
\end{figure*}

In this section, we develop an optimization framework for forecasting MA positions using historical position sequences. First, optimal antenna positions are generated by solving a secrecy rate maximization problem. Subsequently, future MA configurations are forecasted by extracting spatio-temporal patterns from historical position data. The framework of the proposed antenna position optimization and prediction architecture is illustrated in Fig. \ref{fig:process}.

\iffalse 
In this section, the focus is placed on solving the optimization problem for predicting the future positions of the MAs based on their historical position sequences. The dataset is generated by solving a secrecy rate maximization problem to obtain the optimal antenna positions. Then, leveraging the movement patterns of the UAV and MAs, the future antenna positions are predicted based on the historical data. The framework of the proposed antenna position optimization and prediction system is depicted in Fig. \ref{fig:process}.
\fi

\subsection{Dataset construction} 
The PSO algorithm is introduced to solve the secrecy rate maximization problem for antenna position optimization. This approach decomposes the problem into sequential optimization steps, enhancing computational tractability. The PSO iteratively updates MA positions through velocity-driven particle movement, converging to configurations of the secrecy rate maximization. %The pseudo-code of the proposed PSO algorithm is given in algorithm \ref{PSO}.

\iffalse
To address the antenna position optimization problem, which is formulated to maximize the system's secrecy rate, we employ the PSO algorithm. This approach simplifies the problem by breaking it down into a series of manageable steps, making the optimization process more tractable. The PSO algorithm iteratively updates the positions of the MA to find the optimal configuration that maximizes the secrecy rate.
\fi

Initially, $K$ particles are randomly generated with set of positions $\mathbf{q_k^{(0)}}=\{ q_1^{(0)},...,q_k^{(0)},...q_K^{(0)}\}$ and velocities $\mathbf{v}^{(0)} = \{ v_{1}^{(0)},...,v_k^{(0)},...v_K^{(0)}\}$, where each particle
represents a feasible realization of MA's positions. Specially, we get
\begin{equation}
\mathbf{q}_k^{(0)} =
\begin{bmatrix}
\underbrace{x_{k,1}^{(0)},\ y_{k,1}^{(0)},\ z_{k,1}^{(0)}}_{\text{1-th MA}},\
\ldots,\
\underbrace{x_{k,M}^{(0)},\ y_{k,M}^{(0)},z_{k,M}^{(0)}}_{\text{M-th MA}}
\end{bmatrix}^{\textbf{T}},
\end{equation}
where the position $x_{k,m}^{(0)},\ y_{k,m}^{(0)},z_{k,m}^{(0)}$ of the $m$-th antenna is constricted by \eqref{eq:P1-sub2}.

\begin{algorithm}[H] 
	\caption{PSO-based Antenna Position Optimization}  
	\begin{algorithmic}[1]  
		\State \textbf{Initialization:} number of particles $K$, antenna number $M$, maximum iteration number $I_{\max}$, initial positions $\mathbf{q}_k^{(0)}$ and velocities $\mathbf{v}_k^{(0)}$ for all particles 
        \State \textbf{Evaluate:} calculate fitness (secrecy rate) for each particle, set $\mathbf{q}_{k,\text{pbest}} = \mathbf{q}_k^{(0)}$
        \State \textbf{Global best:} set $\mathbf{q}_{\text{gbest}} =  g_{\mathbf{P1}}(\mathbf{q}_{k,\text{pbest}})$
        \For{$i=1$ to $I_{\max}$}
            \For{each particle $k = 1, 2, \dots, K$}
                \State Calculate $\omega^{(i)}$ with Eq. \eqref{eq:omega}
                \State Calculate $\mathbf{v}_k^{(i)}$ with Eq. \eqref{eq:v_update}
                \State Calculate $\mathbf{q}_k^{(i)}$ with Eq. \eqref{eq:q_update}
                \State Evaluate fitness $g_{\mathbf{P1}}(\mathbf{q}_k^{(i)})$
                \If{$g_{\mathbf{P1}}(\mathbf{q}_k^{(i)}) > g_{\mathbf{P1}}(\mathbf{q}_{k,\text{pbest}})$}
                    \State $\mathbf{q}_{k,\text{pbest}} = \mathbf{q}_k^{(i)}$
                \EndIf
                \If{$g_{\mathbf{P1}}(\mathbf{q}_{k,\text{pbest}}) > g_{\mathbf{P1}}(\mathbf{q}_{k,\text{\rm gbest}})$}
                    \State $\mathbf{q}_{\text{gbest}} = \mathbf{q}_{k,\text{pbest}}$
                \EndIf
            \EndFor
        \EndFor
        \State \textbf{Output:} the global best antenna positions $\mathbf{q}_{\text{gbest}}$
	\end{algorithmic}
	\label{PSO}
\end{algorithm}

Each particle updates its velocity and position based on both its individual experience $\mathbf{q}_{\rm pbest}$ (the known local best position) and the global experience $\mathbf{q}_{\rm gbest}$ (the known global best position). The velocity and position of particles at the $(i+1)$-th iteration update based on data at the $i$-th iteration. The velocity update rule of each particle is as follow \cite{Xiao2024Multiuser, PSO1}
\begin{equation}
\begin{aligned}
\mathbf{v}_k^{(i+1)} = &\omega^{(i)} \mathbf{v}_k^{(i)} + c_1 s_1 \left( \mathbf{q}_{k,{\rm pbest}} - \mathbf{q}_k^{(i)} \right) \\
&+ c_2 s_2 \left( \mathbf{q}_{k,{\rm gbest}} - \mathbf{q}_k^{(i)} \right), \label{eq:v_update}
\end{aligned}
\end{equation}
where $\omega$ denotes the inertia weight, which controls how much influence the previous velocity has on the current movement, $c_1$ and $c_2$ are the learning coefficients that determine the relative importance of $\mathbf{q}_{k,{\rm pbest}}$ and $\mathbf{q}_{k,{\rm gbest}}$, $s_1, s_2 \sim \mathcal{U}[0,1]$ are two independent uniformly distributed random variables that introduce stochasticity and improve exploration capability.
In order to balance the exploration capability and convergence speed of the swarm, the inertia weight is adaptively reduced over the iterations \cite{PSO2}
\begin{equation}
    \omega^{i} = \omega_{\max} = \frac{\left( \omega_{\max}- \omega_{\min} \right)i}{I_{\max}}, \label{eq:omega}
\end{equation}
where $\omega_{\max}$ and $\omega_{\min}$ denote the maximum and minimum values of the inertia weight, and ${I_{\max}}$ is the maximum number of iterations. Then, the position of each particle is updated as  
\begin{equation}
\mathbf{q}_k^{(i+1)} = \mathbf{q}_k^{(i)} + \mathbf{v}_k^{(i+1)}. \label{eq:q_update}
\end{equation}
It is worth noting that the optimized positions must satisfy constraint \eqref{eq:P1-sub2}. When an antenna exceeds its predefined spatial bounds, it is projected back onto the boundary. If the displacement between adjacent time slots violates constraint \eqref{eq:P1-sub2}, the movement distance is limited to $d_\mathbf{\max}^{\rm slot}$. The detailed procedure of the proposed PSO-based antenna position optimization is summarized in Algorithm \ref{PSO}.

\begin{itemize}
    \item \textbf{Time complexity:} In this implementation, the computational complexity of the objective function per evaluation can be estimated as $F=O(M L^{\rm nlos}_{t} M_{\rm carol})$, where $M_{\rm carol}$ denotes the Monte Carlo averaging times for random multipath responses.  
    Accordingly, the overall time complexity for optimizing $M$ antenna positions using the PSO algorithm is given by $O( T_{\rm hist} K I_{\max}F) $ \cite{PSO2}.
    \item \textbf{Space complexity:} The space complexity of the proposed PSO-based antenna trajectory optimization algorithm is primarily determined by the number of particles $K$ and the total number of optimization variables $ 3MT_{\rm hist}$, where 3 corresponds to the 3D coordinates. The algorithm needs to store the position, velocity, and personal best position for each particle, as well as the global best position of the swarm. Therefore, the overall space complexity can be expressed as
\begin{align}
  &O\Big(
    \underbrace{3KMT_{\rm hist}}_{\text{position}}
    + \underbrace{3KMT_{\rm hist}}_{\text{velocity}}
    + \underbrace{3KMT_{\rm hist}}_{\scriptsize \text{local optimal}}
    + \underbrace{3MT_{\rm hist}}_{\scriptsize \text{global optimal}}
  \Big) \notag \\
  &\qquad = O(9KMT_{\rm hist} + 3MT_{\rm hist}), \notag
\end{align}
and is reduced to $O(KMT_{\rm hist})$, since both $K$ and $MT_{\rm hist}$ are typically much greater than 1.
\end{itemize}

\subsection{Neural Network Architecture}
\begin{figure*}
\centering
\includegraphics[width=7in ]{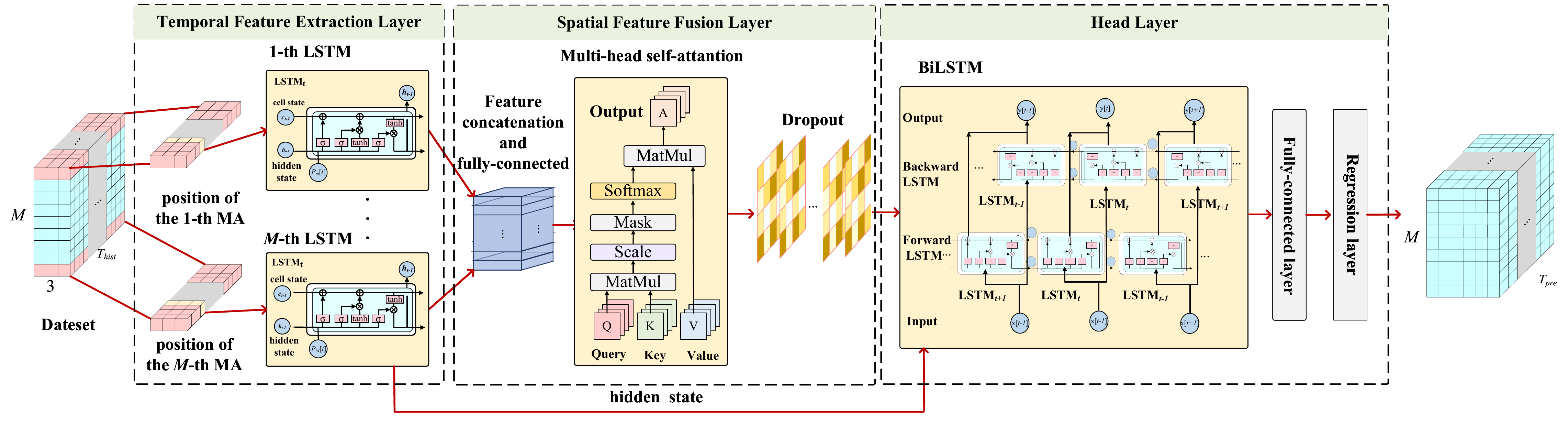}
\caption{\small Neural network framework of the proposed LSTM-Transformer approach}
\label{fig:Neural_Network}
\end{figure*}

Considering the coupling characteristics among multiple antennas, an integrated neural network architecture that combines LSTM and multi-head self-attention mechanisms is proposed to predict the future positions of the antennas through supervised learning \cite{LSTM_trans}. The details of the proposed neural network framework are shown in Fig. \ref{fig:Neural_Network}.

\subsubsection{\textbf{Temporal feature extraction layer}} Each of the $M$ MAs' 3D position trajectories over $T_{\rm{hist}}$ historical slots is modeled independently by a dedicated LSTM network. The $m$-th LSTM processes the $m$-th antenna's position sequence, capturing its unique motion dynamics. As an advanced recurrent neural network variant, LSTM mitigates vanishing/exploding gradient problems in long-sequence processing through its gating mechanism. This architecture regulates temporal information flow via: 1) input gate incorporating new observations, 2) forget gate discarding obsolete information, and 3) output gate controlling state exposure. The cell state maintains long-term contextual memory while hidden states encode short-to-medium-term motion features for subsequent concatenation and fully-connected processing.

 \iffalse
 Based on the dataset containing the 3D positions of $M$ MAs over $T_{\rm hist}$ historical time slots, the temporal evolution of each antenna’s position is independently modeled using a dedicated LSTM network. In this framework, the $m$-th LSTM processes the position sequence of the $m$-th antenna, thereby capturing its unique motion dynamics. As an advanced variant of recurrent neural networks (RNN), the LSTM effectively alleviates the vanishing and exploding gradient problems that often arise when processing long sequential data. This robustness is enabled by its gating mechanism—comprising the input, forget, and output gates—which regulates information flow across time steps. The cell state serves as a long-term memory unit, preserving relevant historical context; the forget gate discards obsolete information, the input gate incorporates new observations, and the output gate determines which parts of the internal state are exposed at the current time step. The resulting hidden states from each LSTM act as temporal feature encoders, representing the short- and medium-term motion characteristics of each antenna for subsequent feature concatenation and fully connected processing.
 \fi

\subsubsection{\textbf{Spatial feature fusion layer}}  
Subsequently, the temporal feature outputs from all $M$ LSTM encoders are concatenated along the feature dimension and transformed via a fully connected layer, while preserving the temporal ordering of the sequences. This design ensures that both the individual motion dynamics of each antenna and the temporal correlations across the antenna array are preserved, thereby providing rich, time-aware feature representations for downstream spatial modeling.

To further capture the spatial dependencies among multiple antennas, a multi-head self-attention mechanism is employed within the spatial feature fusion layer. This mechanism dynamically models the inter-antenna correlations by computing attention weights over the latent representations of all antennas at each time step. As the core component of the Transformer architecture, the multi-head attention mechanism projects the input features into multiple learned subspaces and performs attention computations in parallel. The outputs from each head are then concatenated to form the final spatial representation, enabling the model to capture diverse relational patterns across spatial dimensions while maintaining the continuity of temporal information.
Specifically, the mechanism computes attention scores by linearly transforming the inputs into query ($Q$), key ($K$), and value ($V$) matrices. The attention output is formulated as:
\begin{equation*}
\text{Attention}(Q,K,V) = \text{SoftMax}\left( \frac{QK^{T}}{\sqrt{d_k}} \right) V,
\tag{1}
\end{equation*}
where $d_k$ denotes the dimension of the key vectors, and the $\text{SoftMax}(\cdot)$ function normalizes the attention scores to a probability distribution, thereby highlighting the relative importance of each antenna's contribution.
The “multi-head” design enables the model to concurrently focus on various levels of spatial interactions—for example, strong couplings among nearby antennas and weaker dependencies with distant ones. This enhances the representational capacity of the model by integrating complementary spatial cues from different perspectives.

Moreover, a dropout layer is appended after the attention module to mitigate overfitting by randomly deactivating a subset of neurons during training. This not only improves the generalization performance of the network but also introduces beneficial regularization. Importantly, the temporal structure of the antenna sequence is preserved through step-wise attention refinement, ensuring that the spatial feature extraction does not compromise the continuity of temporal information.

\subsubsection{\textbf{Head layer}}
To enhance the sequential modeling capability, the features generated from the spatial attention mechanism are further enriched with temporal representations derived from the previous LSTM layer. These two sources of information are concatenated to form a comprehensive feature representation: 1) inter-antenna spatial dependencies captured via multi-head self-attention; 2) temporal dynamics embedded in the hidden states of the LSTM. This fused feature is then passed to a bidirectional LSTM (BiLSTM) network to capture bidirectional temporal dependencies more effectively.
Specifically, the BiLSTM consists of two parallel LSTM modules: the forward LSTM processes the sequence from $t=1$ to $t=T_{\rm hist}$, capturing dependencies from past to future; the backward LSTM processes it from $t=T_{\rm hist}$ to $t=1$. At each time step, the outputs of both directions are concatenated, enabling the network to exploit both historical and prospective contexts simultaneously. This dual-view temporal modeling allows the network to better learn complex spatio-temporal patterns across antenna trajectories.
This layered design not only preserves the spatial structure of the antenna array but also strengthens temporal reasoning by leveraging past movement trends and future behavior patterns. As a result, the model achieves more accurate and consistent prediction of antenna positions over time.

Finally, a fully connected layer is employed to project the fused features into the original 3D coordinate space. The network outputs multi-step predictions of 3D positions for all $M$ antennas over $T_{\text{pre}} $ future steps. During training, the network takes historical sequences over $T_{\text{hist}}$ time slots and learns to predict the corresponding $T_{\text{pre}}$ future positions in an autoregressive fashion. The predicted sequence maintains spatial alignment among antennas at each time step and captures their continuous motion dynamics in three dimensions.

\textbf{Model training:}  
The proposed prediction network is trained in a supervised manner using the Adam optimizer, which adaptively adjusts the learning rate for each parameter by maintaining estimates of first and second moments of the gradients. This facilitates faster convergence and better stability, especially in the presence of sparse or noisy gradients often encountered in trajectory prediction tasks.

\section{Simulation Results}\label{sec:simulation}
In this section, numerical simulations are conducted to  demonstrate the robustness of the proposed antenna position optimization and prediction framework. Under the representative scenario, the BS, equipped with $3\times3$ MAs array, transmits signals with the Bob. Each antenna moves within a range of $10\lambda$. Both Bob and Eve fly along the predetermined trajectory. Key simulation parameters are summarized in Table \ref{tab:sim para validation}.

\begin{table}[!htb]  \caption{\small Simulated parameters and values}
\centering
\label{tab:sim para validation}
\begin{tabular}{lll}
\toprule
  Symbol & Meanings &Values\\
\midrule
  $M$          & the number of MAs     & $9$ \\
  $F$          &frequency band of communication & 28GHz \cite{Yujia2024}\\
  $\lambda $   &  beamwave   & $0.0107$m \\
  $\alpha$        & the loss path   & $[2, 4]$ \\
  $H$     & the altitude of BS   & $20$m \\
  $N$     & noise power    & $10^{-5}$W \\
  $P_{\max}$     & the maximum communication power      & 1W \\
  $K$          & the number of particle swarm   & $50$  \\
  $d_{\min}^{\rm MA}$   & the minimum inter-MA distance   & $1/2\lambda$ \cite{Zuo2025} \\
  $\varepsilon$   & accuracy threshold & $0.0005$m   \\ 
  $\eta$ & learning rate & 0.001\\
  \bottomrule
\end{tabular}
\end{table}

\subsection{Baseline frameworks}
To demonstrate the superiority of the proposed framework, which is based on LSTM and Transformer architectures, three baseline methods—NARX, standalone LSTM, and standalone Transformer—are selected for comparison. The details of these baseline methods are described as follows:
\begin{itemize}
\item \textbf{NARX \cite{Yuan2020Machine}:} The NARX is a recurrent dynamic network that models nonlinear time-series relationships by incorporating both past output and input values. It is widely used for nonlinear system identification and time-series prediction tasks.
\item \textbf{LSTM \cite{Liu2021location,Hu2024Location}:} The LSTM network is designed with memory cells and multiplicative gates (input gate, forget gate, and output gate), which enable it to effectively capture long-term dependencies in sequential data. For channel prediction tasks, LSTM can learn and exploit temporal correlations across extended channel state sequences, thereby alleviating the vanishing gradient problem inherent in traditional RNNs.
\item \textbf{Transformer \cite{Xia2024trans,zhou2024Tran}:} The Transformer model is based on a self-attention mechanism and is capable of modeling global dependencies in sequential data without relying on recurrence. This architecture enables parallel processing of input sequences and facilitates the learning of complex temporal relationships, making it particularly effective for sequence predictions in wireless communication systems.
\end{itemize}

\iffalse
\begin{table}[ht]
\centering
\caption{Comparison of Models for Antenna Position Prediction}
\begin{tabular}{|
    >{\centering\arraybackslash}m{1.2cm}|
    >{\centering\arraybackslash}m{3cm}|
    >{\centering\arraybackslash}m{3cm}|}
\hline
\textbf{Model} & \textbf{Advantages} & \textbf{Limitations} \\
\hline
NARX & Simple structure; can model basic nonlinear time-series relationships. & Limited to short-term dependencies; struggles with complex or long-term temporal patterns. \\
\hline
LSTM & Effectively captures long-term temporal dependencies; alleviates vanishing gradient problem. & Limited global context modeling; sequential processing hinders parallelization. \\
\hline
Transformer & Models global dependencies through self-attention; enables parallel processing of sequences. & Requires large datasets and high computational cost; less effective for short-term memory. \\
\hline
\textbf{Proposed} & Integrates bidirectional and long-term temporal modeling with global attention mechanisms; superior performance for antenna position prediction. & Increased model complexity; requires careful tuning and more training data. \\
\hline
\end{tabular}
\label{tab:comparison}
\end{table}
\fi

\subsection{Secrecy rate maximum}

\begin{figure}[!ht]
\centering
\includegraphics[width=2.2in]{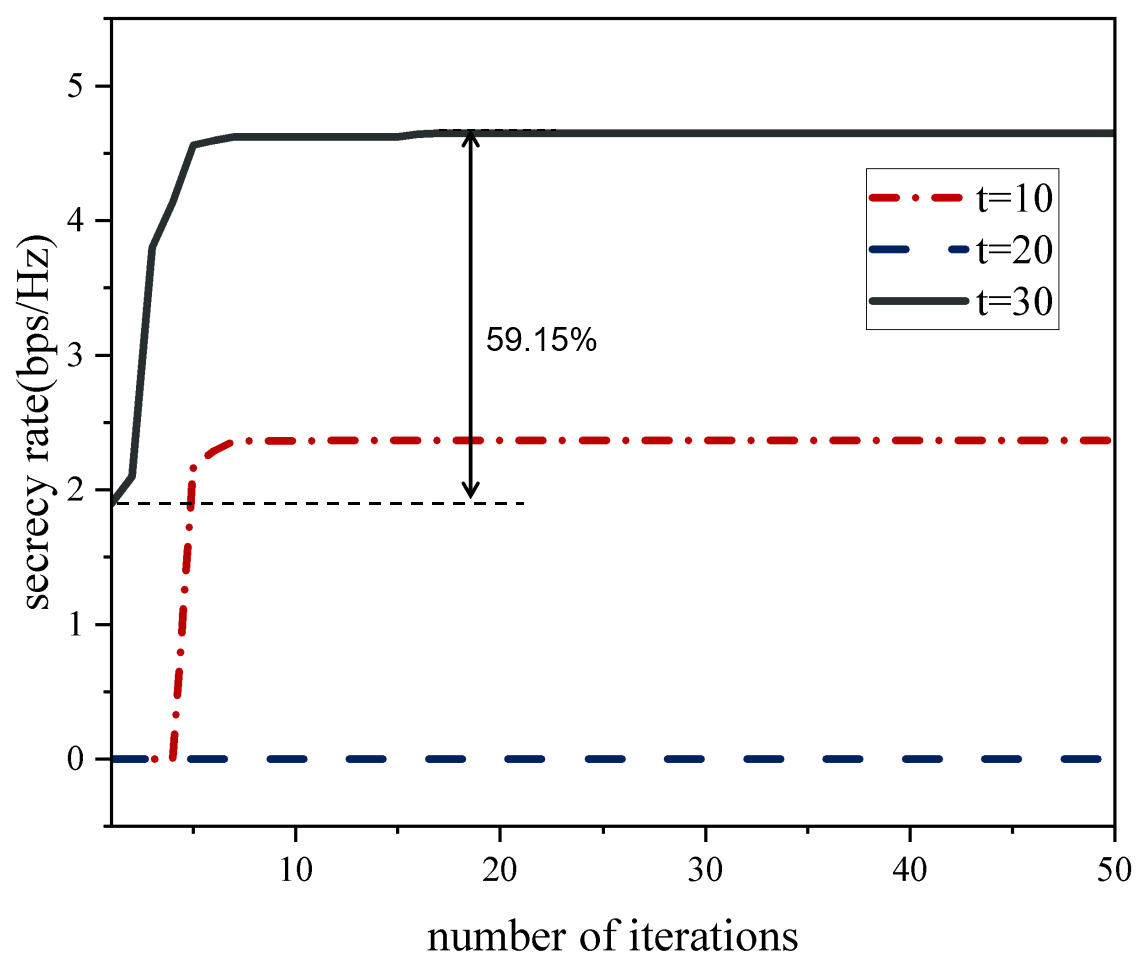}
\caption{\small Secrecy rate  \emph{vs.} the number of iterations}
\label{fig:SR_iter}
\end{figure}

Fig. \ref{fig:SR_iter} illustrates the convergence behavior of the PSO algorithm at different time instances. At time slot $t=10$, the secrecy rate initially remains at zero, but after eight iterations, it rapidly increases to 2.3653 bps/Hz. This result verifies the capability of the PSO algorithm to efficiently search for the optimal antenna positions and substantially enhance the system's physical layer security, even when starting from a suboptimal state. The secrecy rate consistently remains at zero throughout all iterations at $t=20$. This phenomenon is primarily due to the fact that the eavesdropper is located very close to the base station, and the limited mobility of the antennas prevents them from quickly relocating to positions that would ensure positive secrecy performance. For the case of $t=30$, the secrecy rate exhibits a rapid initial increase followed by stabilization, reaching convergence within 20 iterations. The secrecy rate improves from its initial value to a final value of 4.64956 bps/Hz after optimization, highlighting the effectiveness of the PSO algorithm in quickly identifying near-optimal antenna configurations and substantially enhancing physical layer security performance.% under favorable conditions.

\begin{figure}[!ht]
\centering
\includegraphics[width=2.2in]{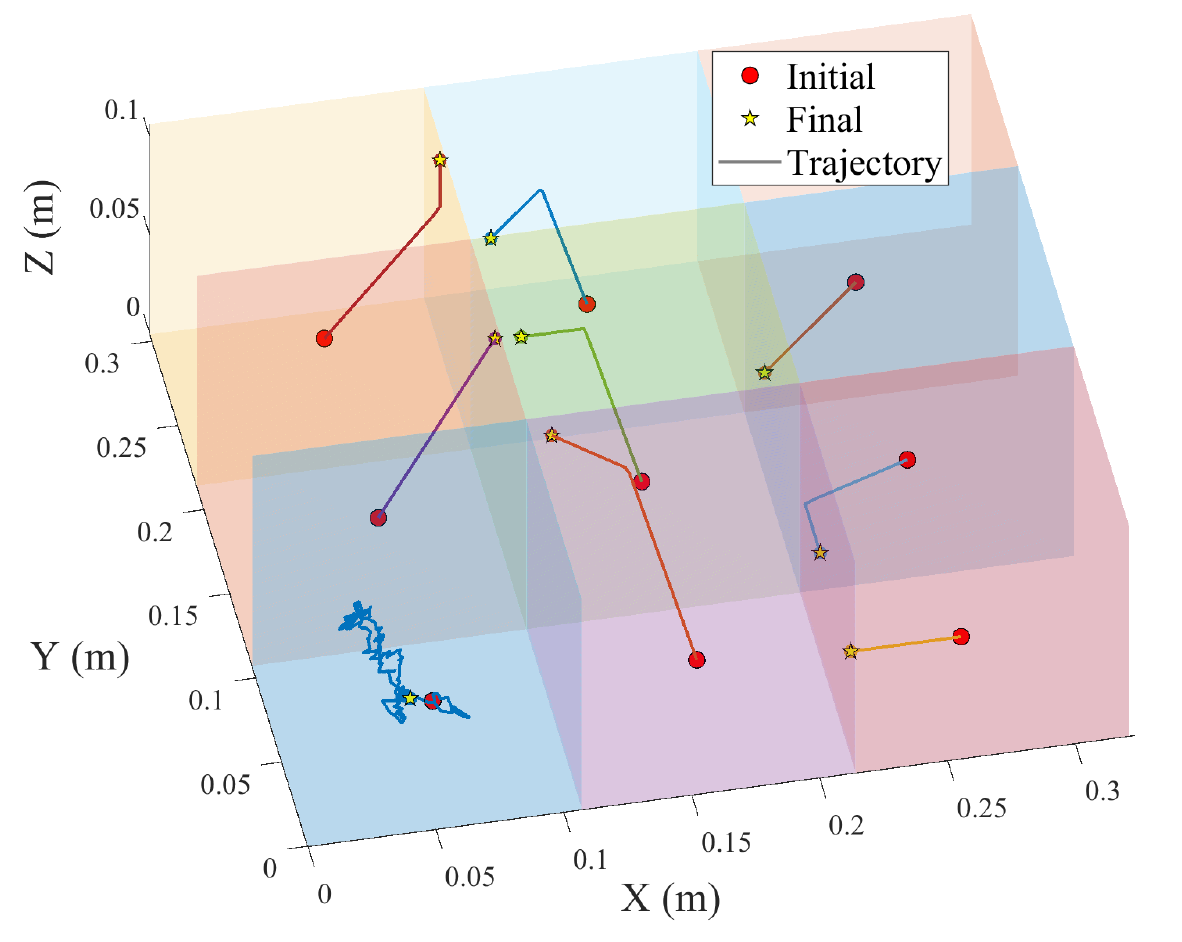}
\caption{\small Moving trajectory of MAs}
\label{fig:MA_position}
\end{figure}

The moving trajectories of MAs are depicted in Fig. \ref{fig:MA_position}, where all antennas initially move collectively toward the optimal direction to enhance system performance. As time progresses, spatial constraints become more pronounced, causing antennas 2 through 9 to quickly reach their respective boundary positions, beyond which further movement is restricted; consequently, these antennas exhibit oscillatory behavior within these confined regions, while the first antenna persistently adjusts its position in a more dynamic manner to sustain overall system optimality. This behavior demonstrates that the proposed algorithm not only enables efficient initial convergence, but also maintains adaptive control of key antenna elements to respond to environmental changes and system constraints, thereby maximizing the secrecy rate in a dynamic wireless environment.

\begin{figure*}[!ht]
    \centering
    \begin{subfigure}{0.32\textwidth}
        \centering
        \includegraphics[width=0.95\linewidth]{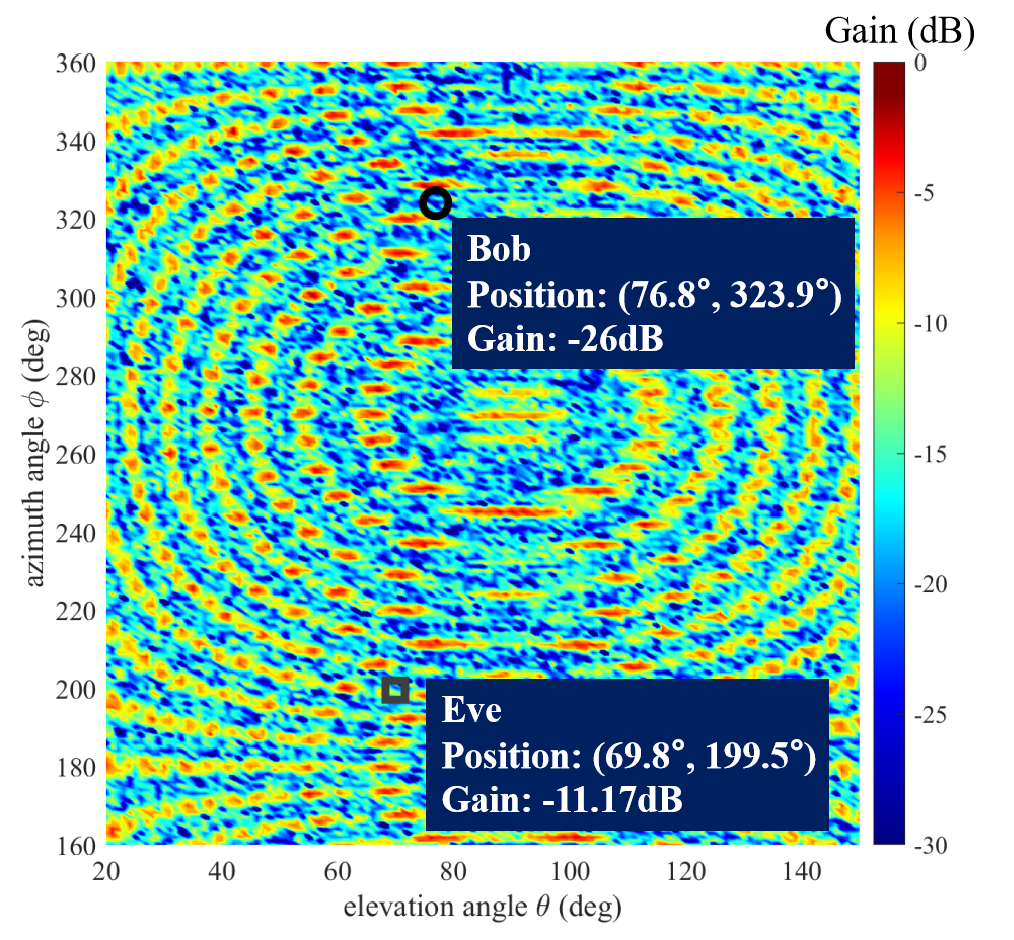}
        \caption{\small Previous optimization}
        \label{fig:gain_fix}
    \end{subfigure}
    \begin{subfigure}{0.32\textwidth}
        \centering
        \includegraphics[width=\linewidth]{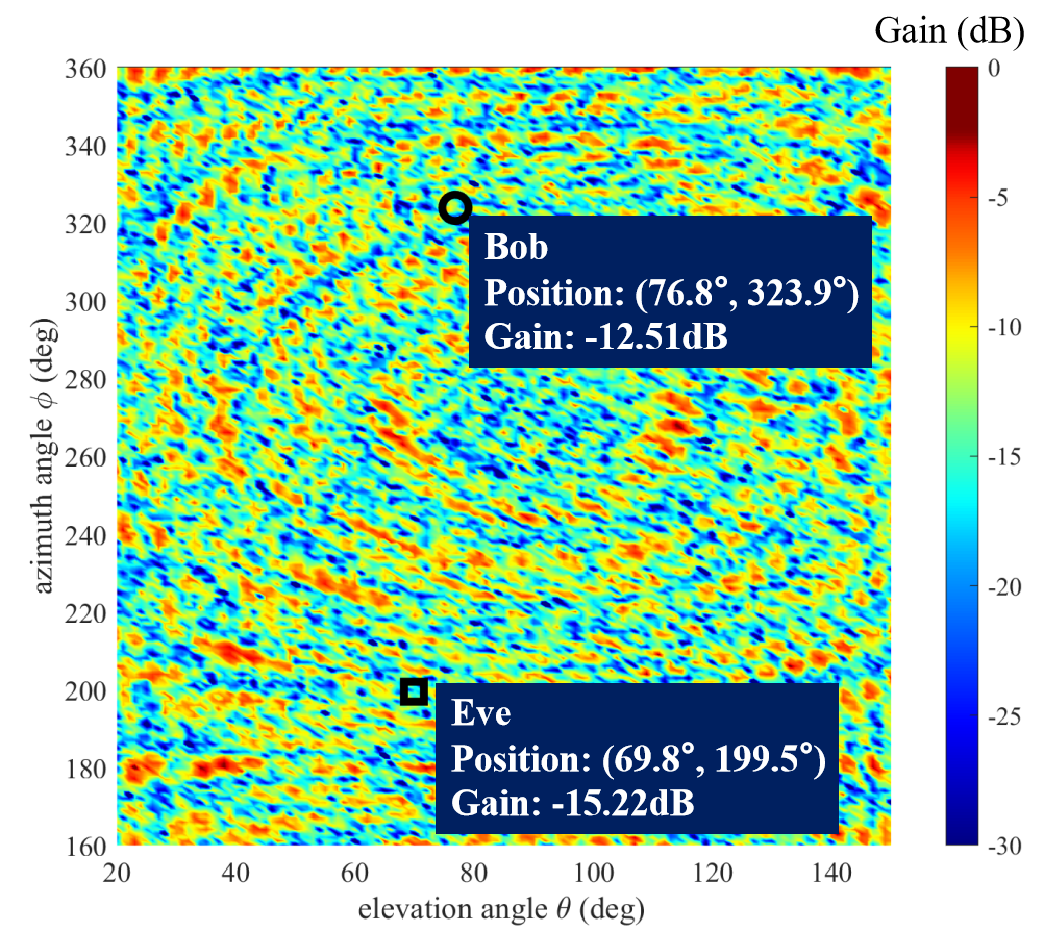}
        \caption{\small MA positions optimization}
        \label{fig:gain_MA}
    \end{subfigure}
    \begin{subfigure}{0.32\textwidth}
        \centering
        \includegraphics[width=\linewidth]{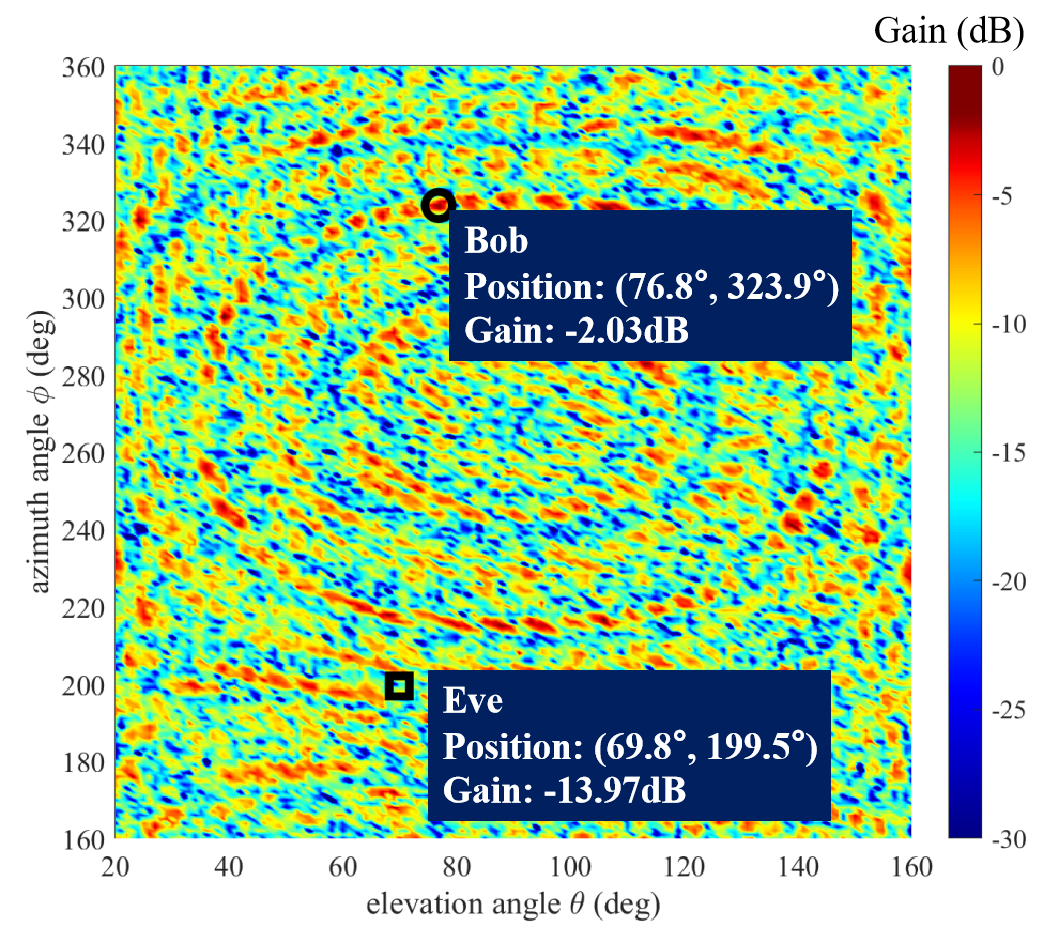}
        \caption{\small Joint optimization of MA position and BF}
        \label{fig:gain_MA_BF}
    \end{subfigure}
    \caption{Array pattern gain (antenna movement range $10\lambda$)}
    \label{fig:gain}
\end{figure*}

\begin{figure*}[!ht]
    \centering
    \begin{subfigure}{0.32\textwidth}
        \centering
        \includegraphics[width=0.95\linewidth]{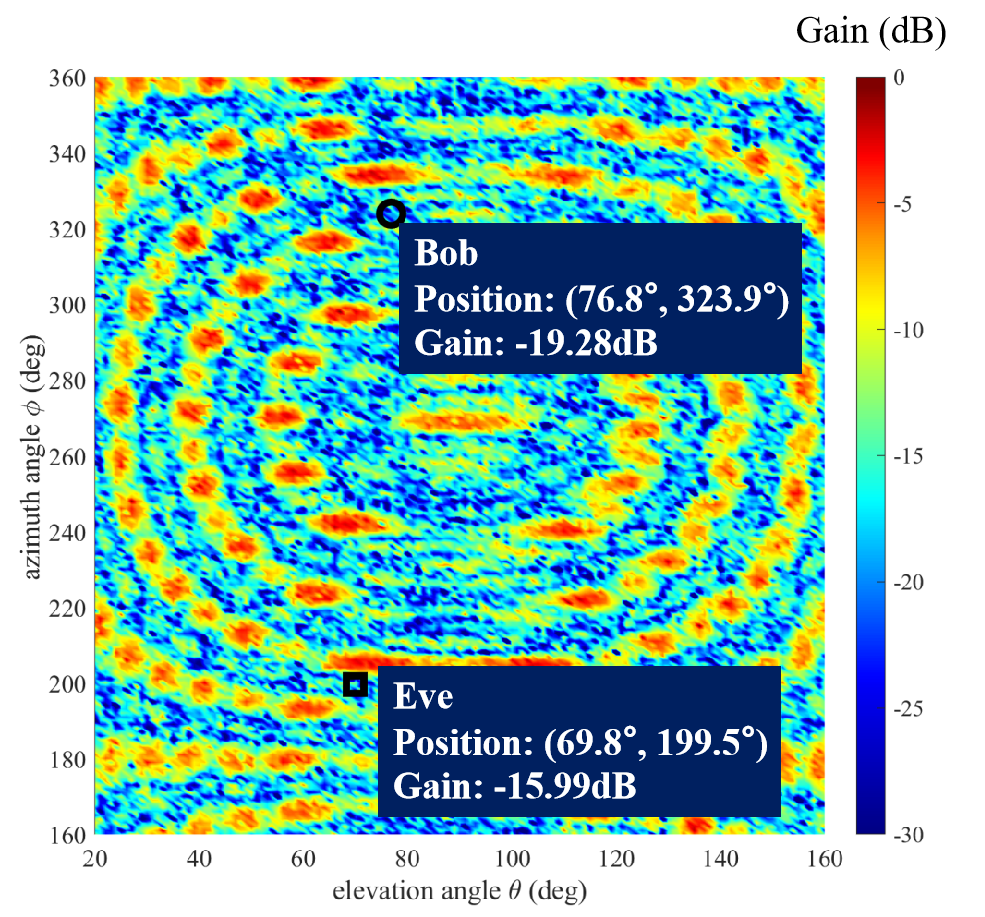}
        \caption{\small Previous optimization}
        \label{fig:gain_fix2}
    \end{subfigure}
    \begin{subfigure}{0.32\textwidth}
        \centering
        \includegraphics[width=\linewidth]{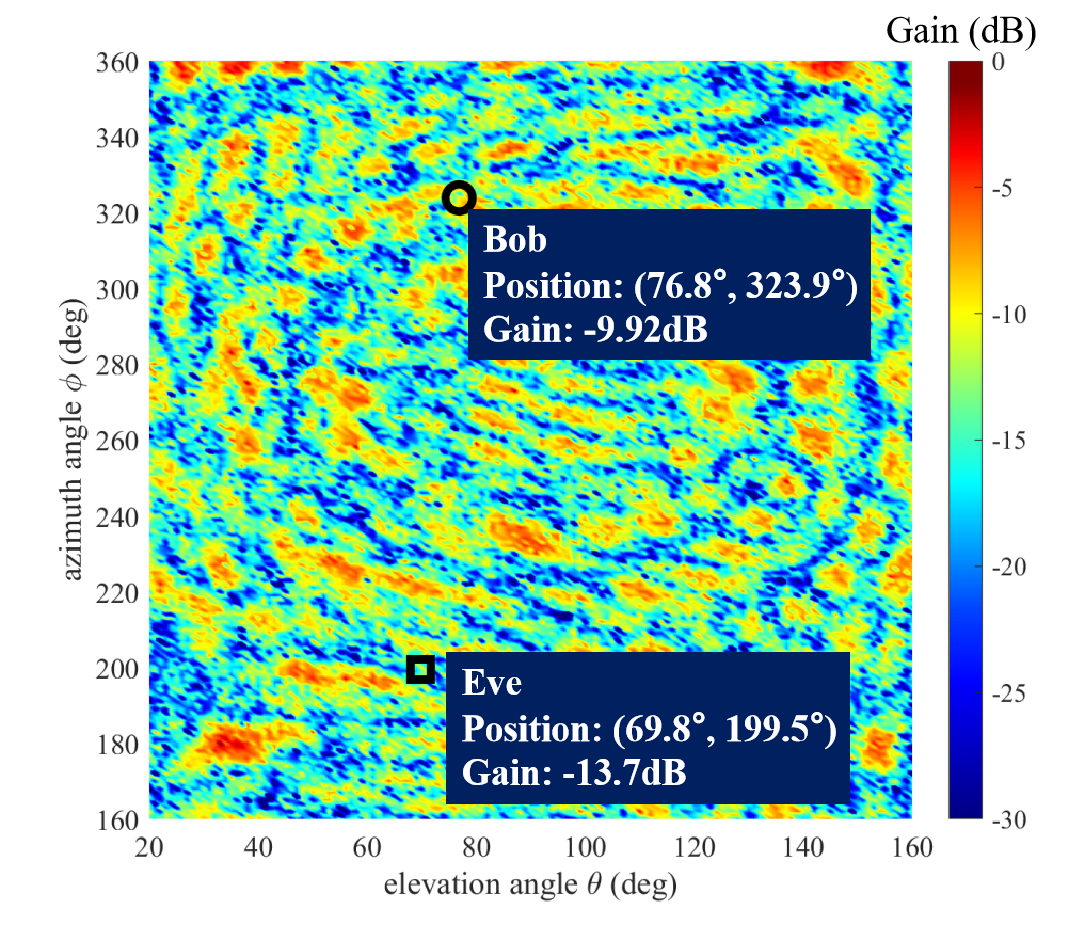}
        \caption{\small MA positions optimization}
        \label{fig:gain_MA2}
    \end{subfigure}
    \begin{subfigure}{0.32\textwidth}
        \centering
        \includegraphics[width=0.95\linewidth]{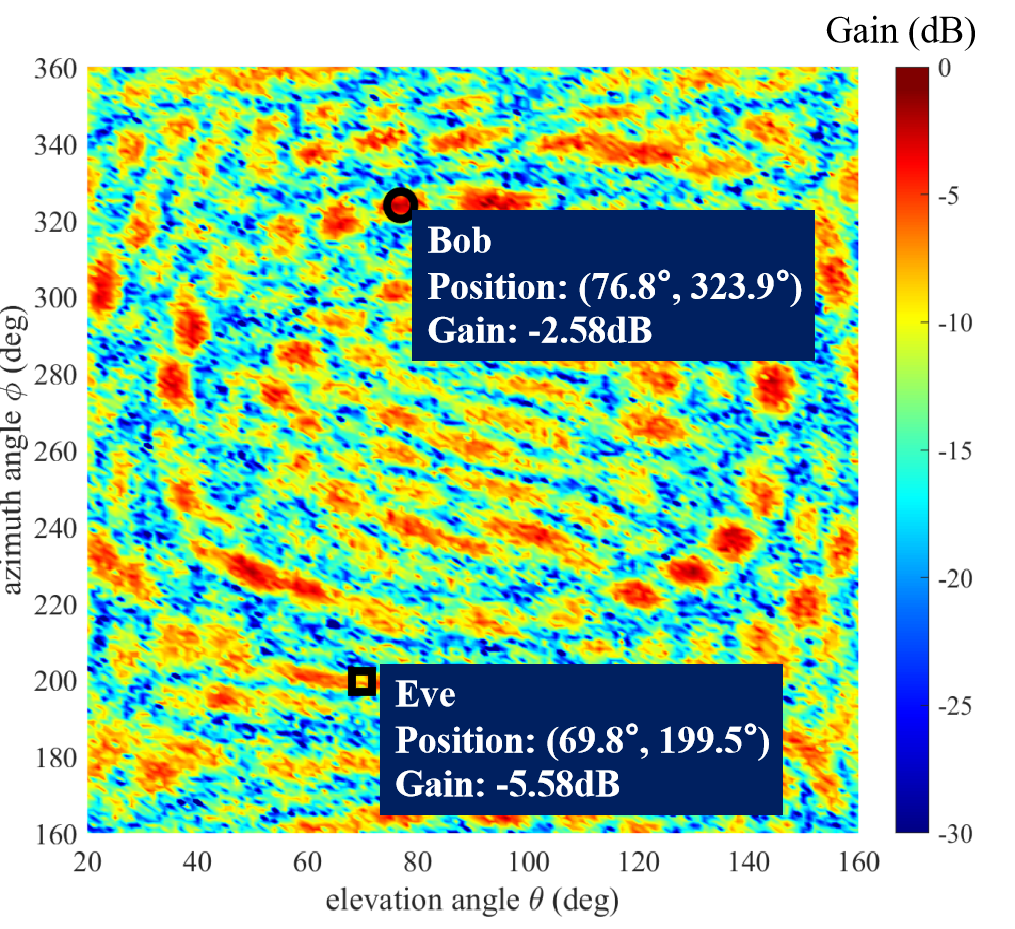}
        \caption{\small Joint optimization of MA position and BF}
        \label{fig:gain_MA_BF2}
    \end{subfigure}
     \caption{\small Array pattern gain (with antenna movement range of $5\lambda$)}
    \label{fig:gain2}
\end{figure*}

Fig. \ref{fig:gain} and Fig. \ref{fig:gain2} compare the array pattern gain of different optimization schemes under antenna movement ranges of $10\lambda$ and $5\lambda$, respectively. For the fixed antenna scenario, the gain in Bob's direction is much lower than that in Eve's direction. After optimizing the MA positions, the gain towards Bob becomes superior to that towards Eve. With further beamforming optimization, the main lobe gain in Bob's direction is further enhanced. 
In addition, a larger antenna movement range provides the array with greater spatial flexibility and enables higher achievable gain in the target direction. However, when the antenna spacing increases to $10\lambda$, more grating lobes are generated, which reduces the directivity and energy efficiency of the array, and makes it more difficult to uniquely focus energy in the desired direction. Therefore, it is necessary to appropriately set the movable range of the MA to achieve a favorable trade-off between flexibility and beam directivity.

\subsection{Effectiveness evaluation of the proposed framework}
Next, the effectiveness of our proposed antenna position prediction framework is verified through a comprehensive performance comparison with three benchmark models: NARX, LSTM, and Transformer. 

\begin{figure}[!ht]
\centering
\includegraphics[width=2.2in]{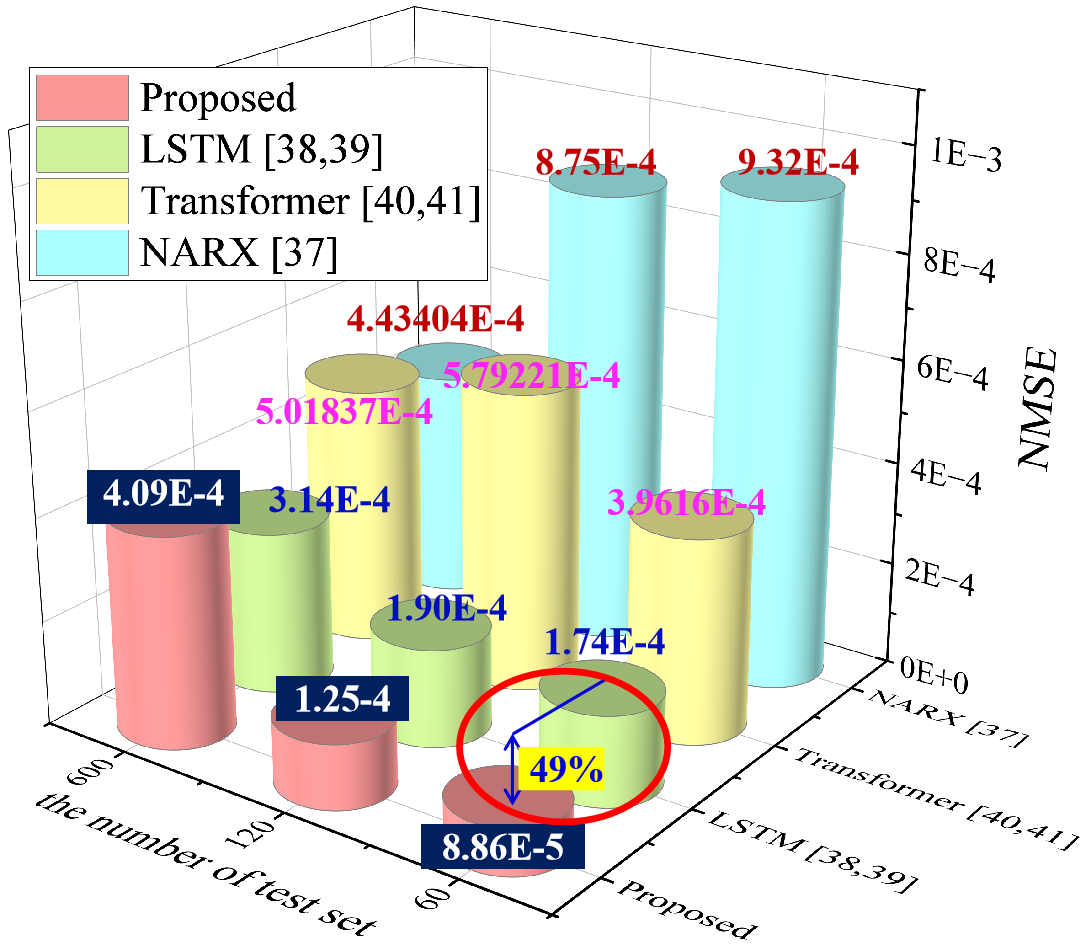}
\caption{\small NMSE   \emph{vs.} methods with different numbers of test sets }
\label{fig:NMSE}
\end{figure}

The NMSE performance of the proposed approach with baselines among the number of test set are compared in Fig. \ref{fig:NMSE}. It can be observed that all of the evaluated methods exhibit consistently low NMSE, with values remaining below 0.001. This can be attributed to the limited spatial range (within $10\lambda$) of antenna movement in the dataset, which reduces prediction difficulty. Nevertheless, subtle variations in antenna positions can induce significant fluctuations in channel conditions, thereby exerting a pronounced impact on communication performance. Notably, the proposed method significantly outperforms the baseline methods in terms of NMSE, owing to its superior ability to jointly learn temporal evolution and spatial dependencies among antenna positions. In addition, NMSE improves as the expansion of the test set grows, which can be attributed to the compounding of prediction errors over extended time horizons, as well as the increased complexity and variability of antenna motion patterns covered in longer test sequences. Under a prediction set size of 60, the proposed model outperforms the baseline frameworks with at least a 49$\%$ reduction in NMSE. Interestingly, the NARX model exhibits relatively better performance when the test set size reaches 600. This could be due to its strong memory capability in capturing repeated historical patterns in relatively constrained mobility settings.

\begin{figure}[!ht]
\centering
\includegraphics[width=2.5in]{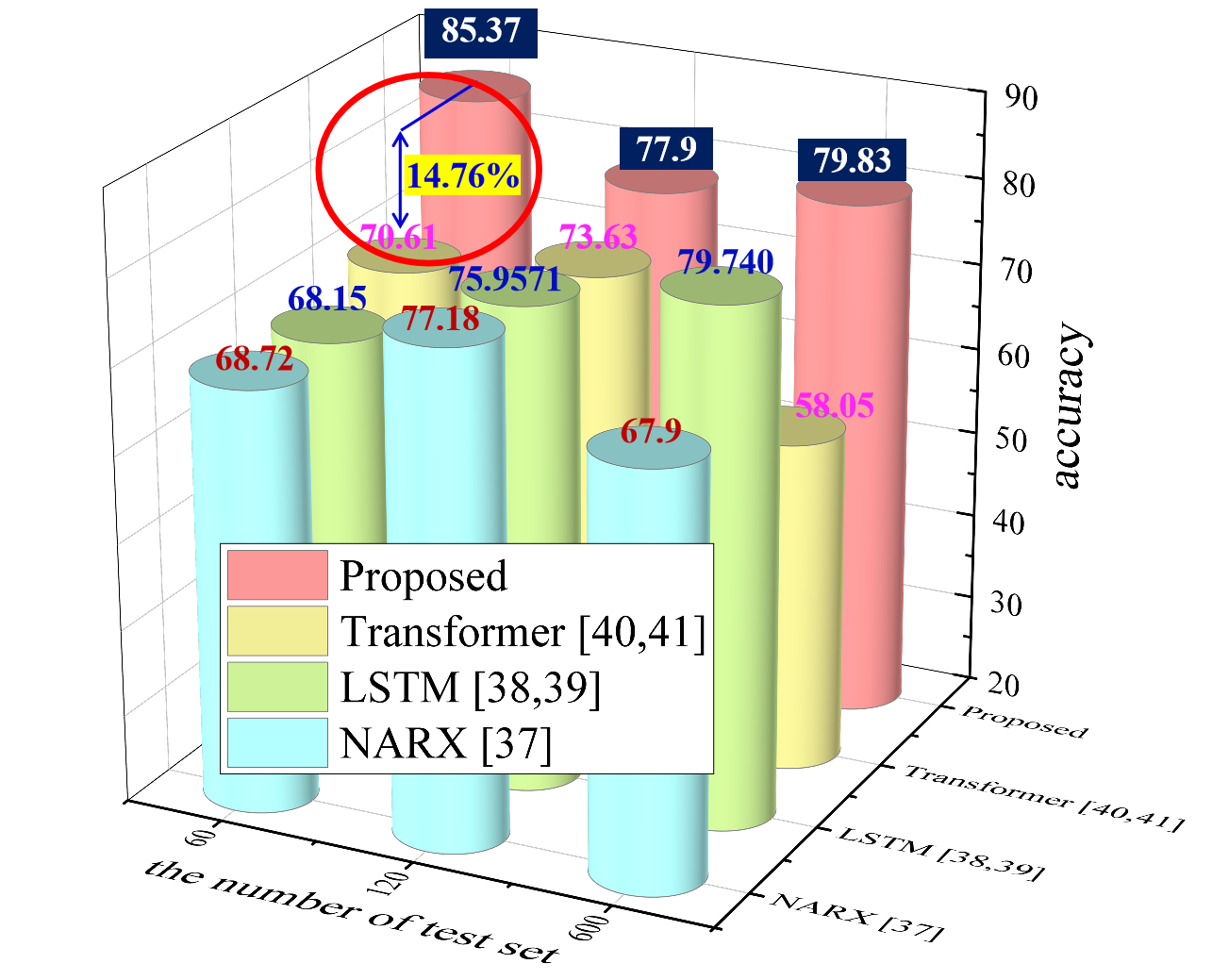}
\caption{\small Accuracy  \emph{vs.} framework with different numbers of test sets }
\label{fig:accuracy}
\end{figure}

Fig.~\ref{fig:accuracy} illustrates the prediction accuracy of different models under a threshold of 0.0005m. It can be observed that the proposed framework consistently outperforms all baseline methods across various test set sizes.For a test set size of 60, the proposed scheme delivers a minimum accuracy improvement of $14.76\%$ over the baseline frameworks, demonstrating its substantial performance advantage. This substantial improvement further corroborates the trend observed in Fig. \ref{fig:NMSE}, where our model demonstrates superior robustness and generalization in predicting antenna positions.

\begin{figure}[!ht]
\centering
\includegraphics[width=2.5in]{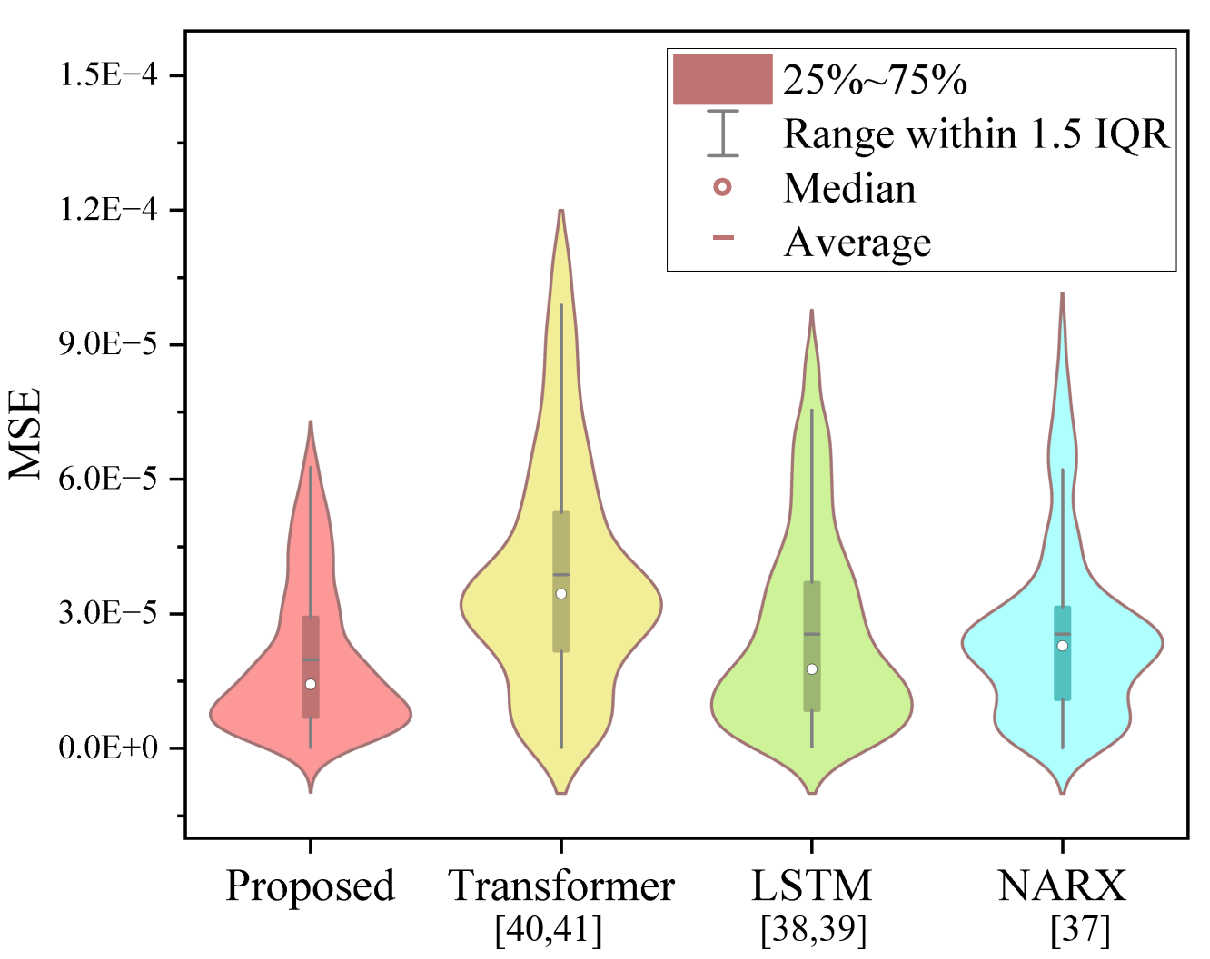}
\caption{\small MSE  \emph{vs.} different frameworks}
\label{fig:MSE}
\end{figure}

The MSE distribution for different models is visualized in Fig. \ref{fig:MSE}, where each colored region reflects the error dispersion of a method. The central box indicates that the interquartile range varies between 25\% and 75\%, the extended vertical line illustrates the span within 1.5 times the interquartile range (IQR). The proposed method achieves a lower MSE, with a compact distribution concentrated around smaller error values. In contrast, the Transformer model yields a wider and higher error spread, indicating less stability and accuracy. LSTM and NARX demonstrate moderate performance but still lag behind the proposed framework in terms of both median error and variance. These results highlight the superior precision and robustness of the proposed framework in position prediction.

\subsection{Secrecy performance evaluations}
To evaluate the communication effectiveness among different models, secrecy rate is adopted as the primary performance metric under various parameter settings. The results provide a comprehensive comparison of the proposed scheme with baseline models such as LSTM, Transformer, and NARX, demonstrating its advantages in terms of security and robustness under different communication scenarios.

\begin{figure}[!ht]
\centering
\includegraphics[width=2.2in]{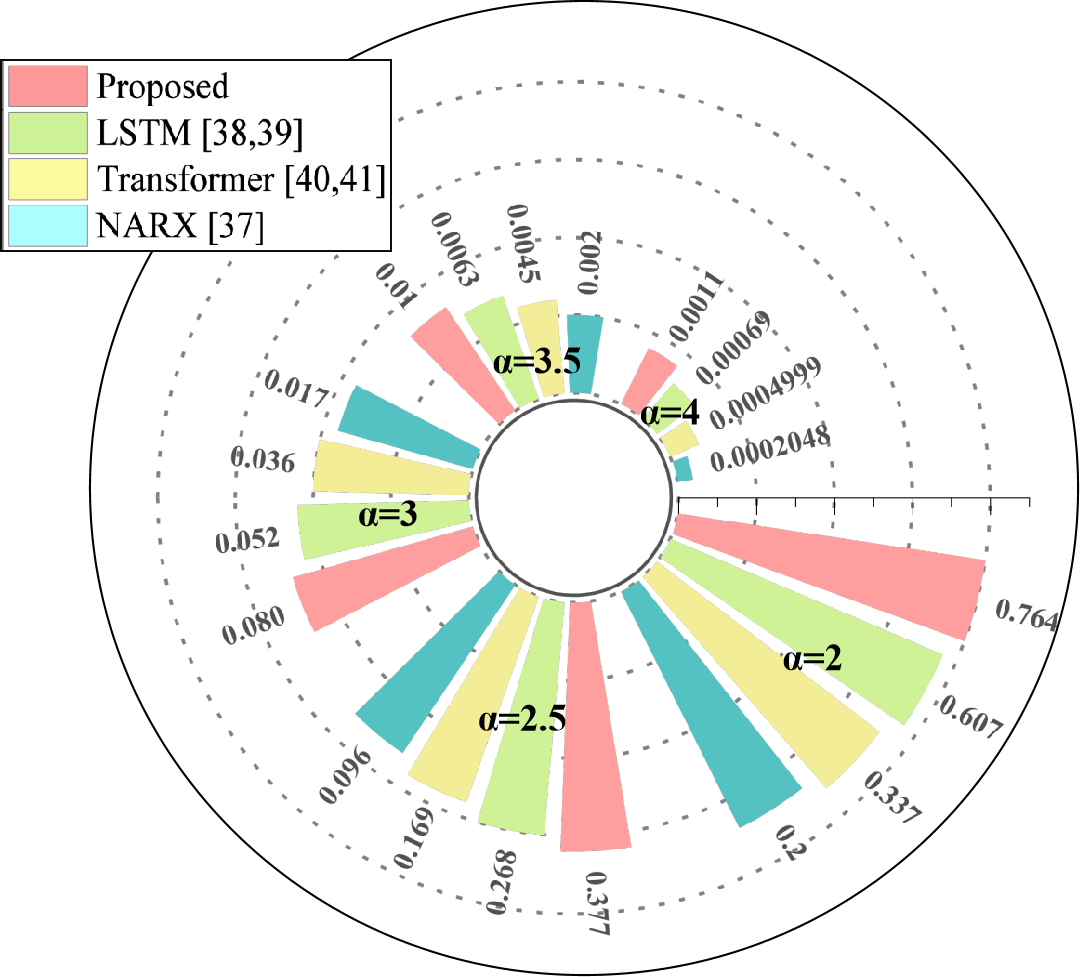}
\caption{\small Secrecy rate  \emph{vs.} path loss }
\label{fig:SR_loss}
\end{figure}

Fig. \ref{fig:SR_loss} demonstrates the secrecy rate of different models as the path loss exponent varies. All models experience a consistent decline in secrecy rate with increasing path loss exponent, which can be attributed to the fact that higher path loss result in more severe signal attenuation, reducing the quality of the received signal. While the proposed model consistently outperforms the other schemes, maintaining a higher secrecy rate and demonstrating superior robustness under increasingly adverse propagation conditions. 

\begin{figure}[!ht]
\centering
\includegraphics[width=3in]{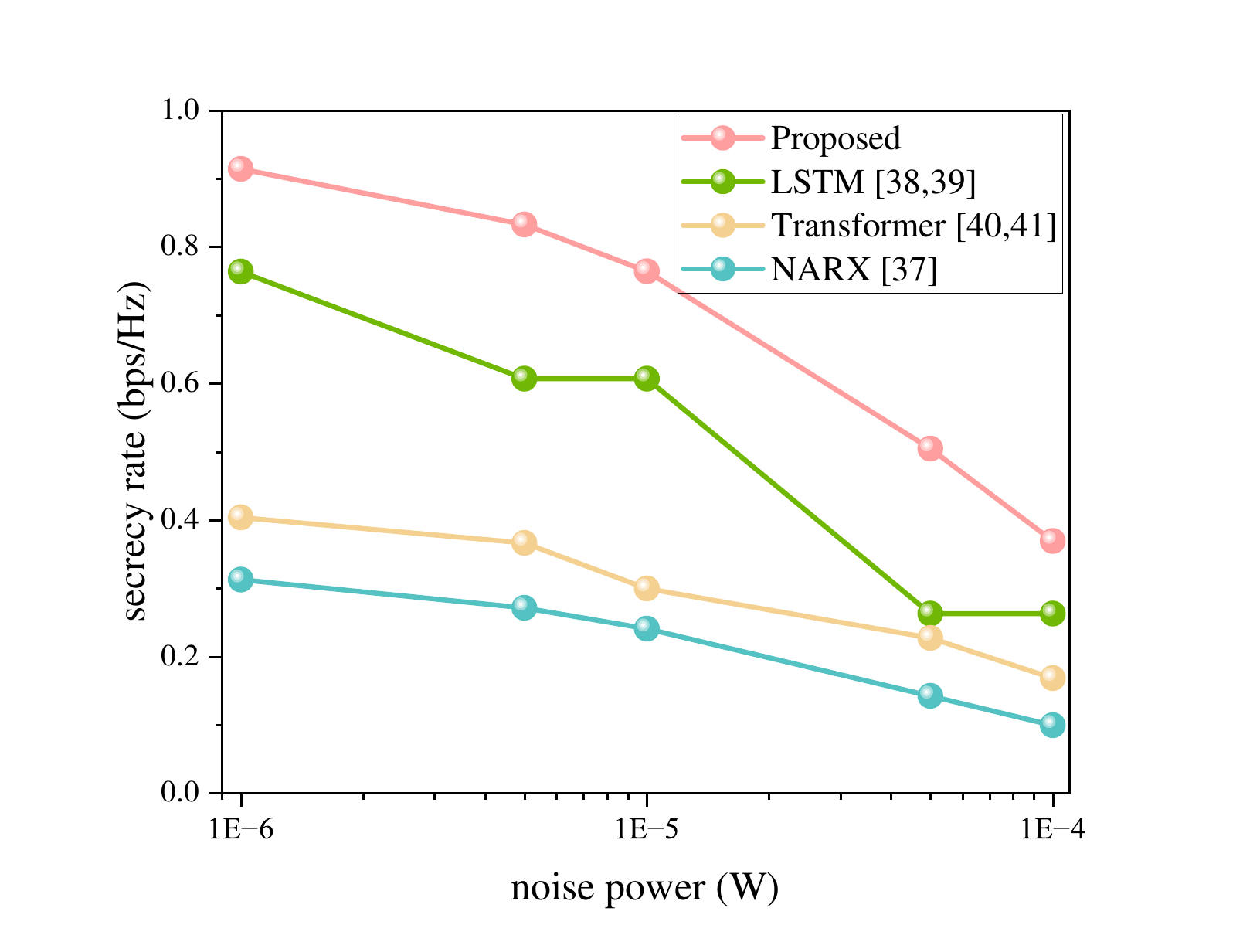}
\caption{\small Secrecy rate  \emph{vs.} noise power }
\label{fig:SR_Noisepower}
\end{figure}

The impact of noise power on communication performance is illustrated in Fig. \ref{fig:SR_Noisepower}. It can be observed that the secrecy rate of all models decreases as noise power increases, since higher noise levels lead to a reduced SNR, which in turn degrades the quality of the received signal and increases the difficulty of maintaining secure and reliable communications. Corresponding to the results shown in Fig. \ref{fig:SR_loss}, the proposed model consistently surpasses the baseline schemes, indicating its superior robustness and enhanced capability to maintain secure communication under adverse channel conditions.

The influence of communication power on secrecy rate achieved by different models is shown in Fig. \ref{fig:SR_loss}. All schemes demonstrate a high level of secrecy rate with increased communication power, which enhancement is principally due to the fact that increased communication power amplifies the legitimate signal at the receiver side, thereby boosting the SNR. The proposed scheme has consistently demonstrated superior performance in comparison to alternative models, a discrepancy that becomes increasingly evident as the communication power increases. This further corroborates the previous findings, demonstrating the remarkable capability of the proposed method.

\begin{figure}[!ht]
\centering
\includegraphics[width=3in]{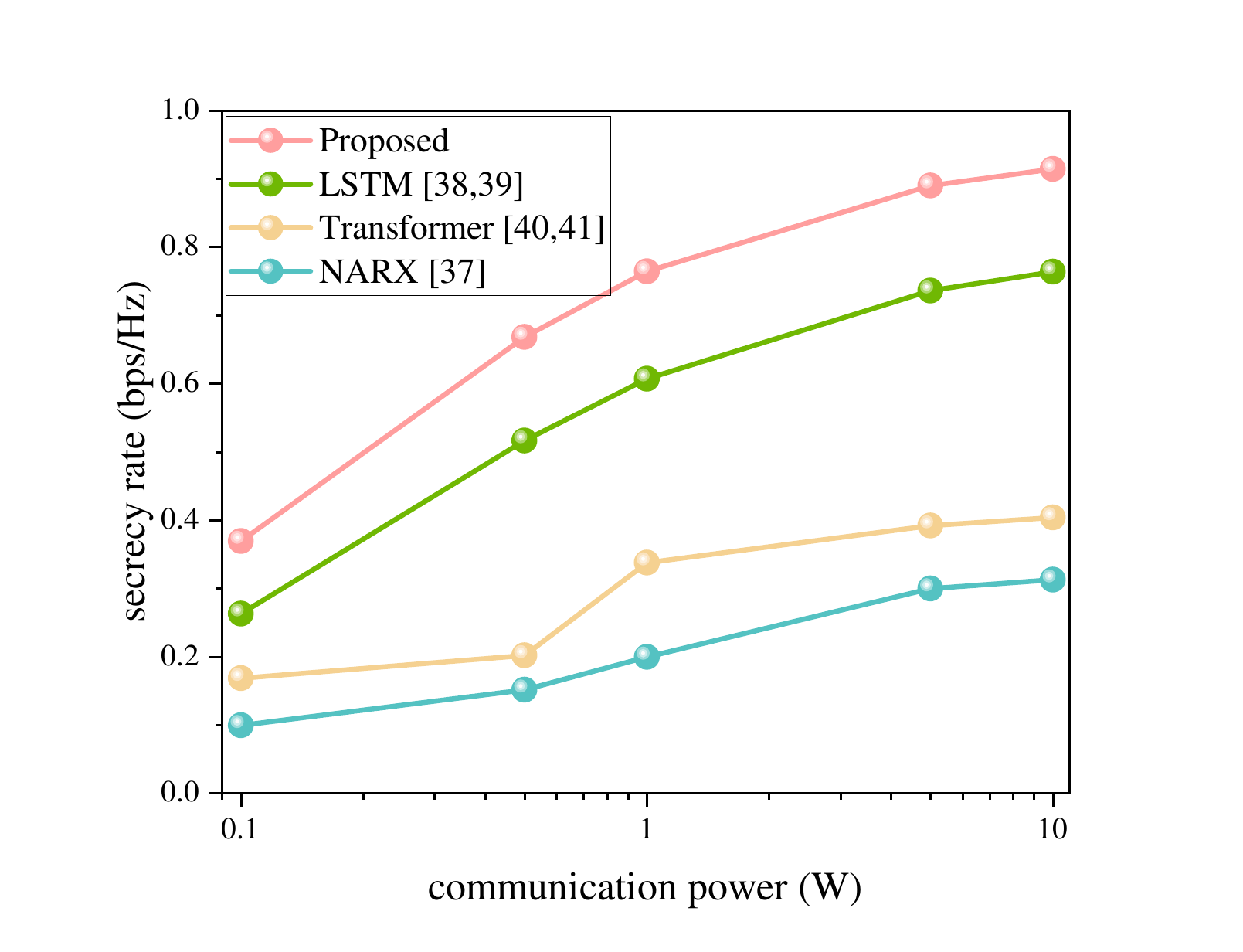}
\caption{\small Secrecy rate  \emph{vs.} communication power }
\label{fig:SR_Compower}
\end{figure}

\begin{figure}[!ht]
\centering
\includegraphics[width=2.5in]{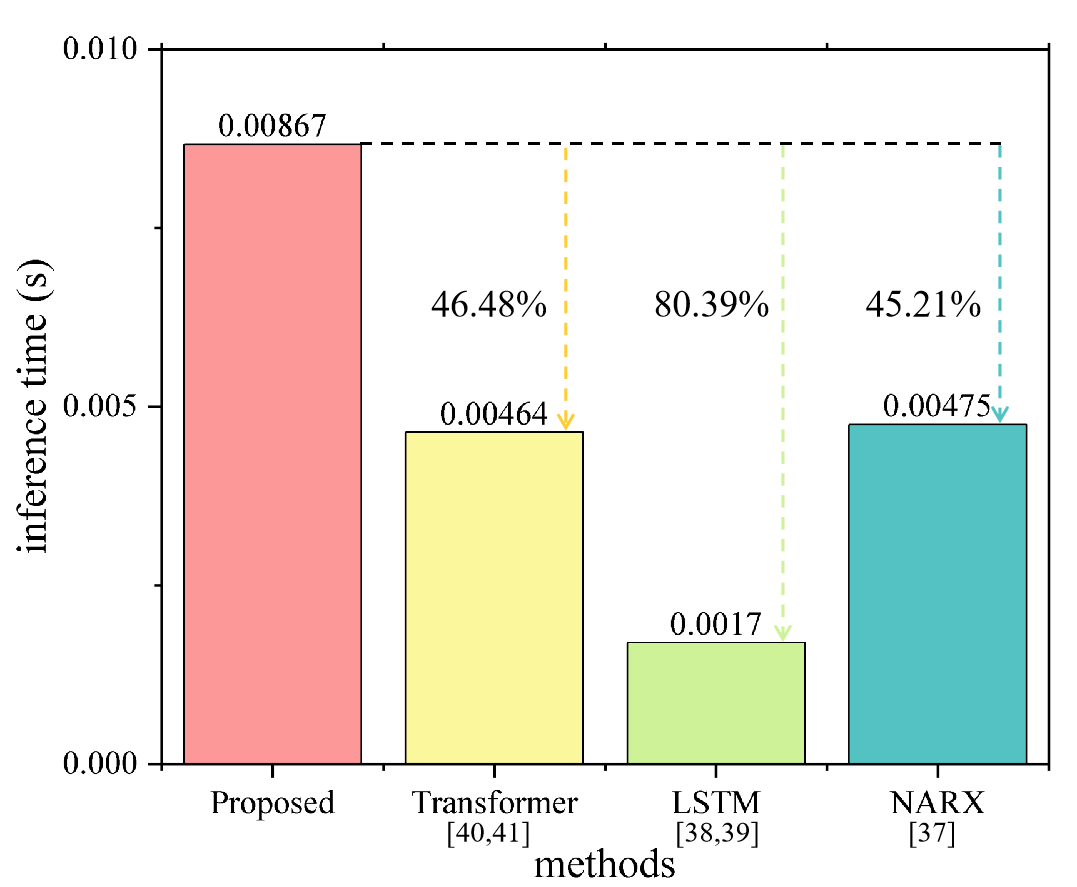}
\caption{\small Inference time \emph{vs.} different frameworks }
\label{fig:time}
\end{figure}

Fig. \ref{fig:time}  demonstrated the inference time among the proposed model and baseline models. Compared with other methods, the proposed approach exhibits a higher inference time, which is mainly attributed to the increased model complexity resulting from the integration of both LSTM and Transformer architectures to enhance prediction accuracy. 
Despite the increased computational cost, the inference time remains within an acceptable range for practical deployment.
On one hand, the inference time required is only 8.67ms, which is significantly lower than the typical time durations (typically 0.5s-1s \cite{time1,time2}) widely adopted in UAV communication and trajectory prediction applications. On the other hand, in terms of absolute latency, the inference time is also much lower than the stringent latency requirements specified in current standards (e.g., 50 ms in 3GPP specifications for command and control \cite{time3}). Moreover, despite the increased computational cost, the inference time remains within an acceptable range for practical deployment, especially considering the substantial improvements in prediction accuracy and robustness achieved by the proposed method. Therefore, the inference time of the proposed method remains well within the acceptable range for practical applications.

\begin{figure}[!ht]
\centering
\includegraphics[width=2.5in]{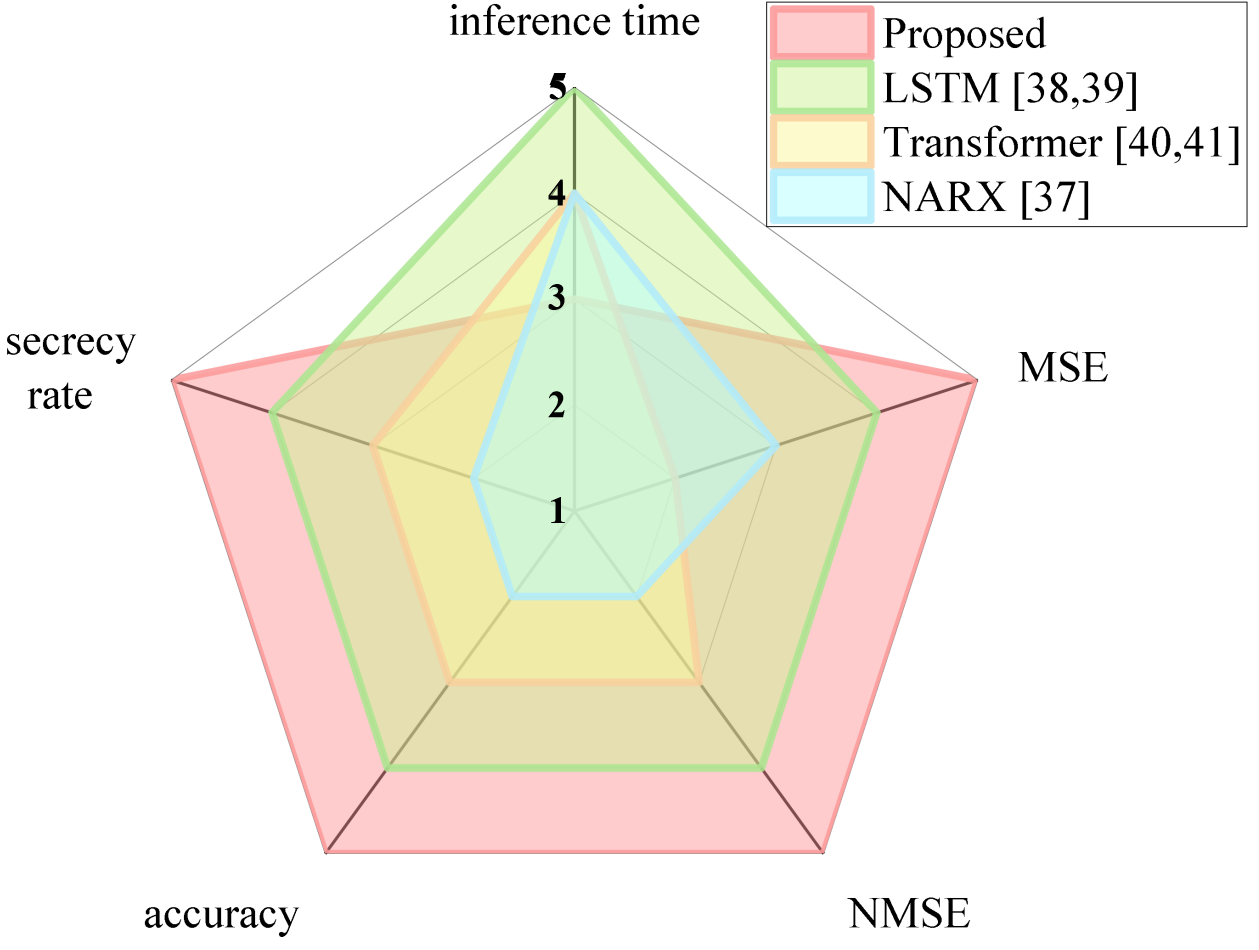}
\caption{\small Comprehensive performance comparison}
\label{fig:radar}
\end{figure}

Finally, in Fig. \ref{fig:radar}, a comprehensive performance comparison is presented for all methods across key metrics, including inference time, MSE, NMSE, accuracy, and secrecy rate. Although a certain degree of inference time is sacrificed by the proposed scheme due to its higher model complexity, superior performance is demonstrated in both position prediction accuracy and physical-layer security. This result highlights the effectiveness and practical value of the proposed approach in secure and intelligent antenna position prediction scenarios.

\section{Conclusion}\label{sec:conclusion}
In this paper, we presented a predictive antenna control framework that dynamically adapts to UAV mobility while preserving robust communication performance. By formulating a secrecy rate maximization problem, we derived optimal antenna positions under dynamic channel conditions using an efficient PSO solver. To overcome mechanical repositioning latency, a hybrid LSTM-Transformer network predicted future antenna positions through spatio-temporal trajectory learning. This anticipatory adjustment enabled proactive antenna relocation, effectively mitigating mobility-induced performance degradation. Extensive simulations confirmed the proposed framework's superiority: it achieves significant prediction accuracy gains and reduces positioning errors with 49\% at least compared to benchmarks, while maintaining computational latency within practical operational thresholds despite moderately increased model complexity.

\iffalse
In this paper, we propose a predictive antenna control framework that adapts to UAV mobility while maintaining robust communication performance. Specifically, a secrecy rate maximization problem is formulated to identify the optimal antenna positions under dynamic channel conditions, which is efficiently solved by a PSO algorithm. To address the latency in mechanical repositioning and ensure real-time adaptability, we integrate LSTM and Transformer architectures to design a hybrid prediction network, which forecasts future antenna positions based on historical trajectories. This proactive adjustment enables the antennas to relocate in advance, thereby mitigating the performance degradation caused by mobility-induced delays. Extensive simulations demonstrate that although the proposed model incurs slightly higher inference time due to the increased model complexity, it significantly outperforms baseline methods in terms of prediction accuracy and positioning error, while keeping the computational latency within a practically acceptable range.
\fi

\bibliographystyle{IEEEtran}
\bibliography{ref}
\end{document}